\documentclass[openacc]{rstransa}

\usepackage{epsfig}
\usepackage{cuted}
\usepackage{amsmath}
\usepackage{amssymb}
\usepackage{float}
\usepackage{xcolor}
\usepackage[normalem]{ulem}
\usepackage{multirow}
\usepackage{multicol}
\usepackage{subcaption}
\usepackage{verbatim}
\usepackage{MnSymbol}
\usepackage{bm}
\usepackage{mathtools}
\usepackage[export]{adjustbox}
\usepackage{stackengine}
\usepackage{amsfonts}
\usepackage{amsthm}
\usepackage{graphicx}
\usepackage{times}
\usepackage{helvet}
\usepackage{courier}
\usepackage{url}
\usepackage{babel}
\usepackage{dcolumn}
\usepackage{psfrag}
\usepackage{setspace}

\usepackage[a4paper,margin=1.475in]{geometry}

\definecolor{b}{rgb}{0, 0.4470, 0.7410}
\definecolor{r}{rgb}{1, 0.3010, 0.3010}
\definecolor{o}{rgb}{0.8500, 0.3250, 0.0980}
\definecolor{y}{rgb}{0.9290, 0.6940, 0.1250}
\definecolor{g}{rgb}{0.4660, 0.6740, 0.1880}
\definecolor{p}{rgb}{0.4940, 0.1840, 0.5560}

\jname{rsta}
\Journal{Phil. Trans. R. Soc}

\begin{document}

\title{Bio-inspired periodic panels optimised for acoustic insulation}

\author{Vin\'icius F. Dal Poggetto$^1$, Nicola M. Pugno$^{1,2}$, Jos\'e Roberto de F. Arruda$^3$}

\address{$^{1}$Laboratory for Bio-inspired, Bionic, Nano, Meta Materials \& Mechanics, Department of Civil, Environmental and Mechanical Engineering, University of Trento, 38123 Trento, Italy\\
$^{2}$School of Engineering and Materials Science, Queen Mary University of London, Mile End Road, London E1 4NS, United Kingdom\\
$^{3}$Department of Computational Mechanics, School of Mechanical Engineering, University of Campinas, Brazil}

\subject{Mechanical Engineering}
\keywords{Sound transmission loss, Bio-inspired structure, Periodic plate, Plane wave expansion, Vibrations}

\corres{
Vin\'icius F. Dal Poggetto\\ \email{v.fonsecadalpoggetto@unitn.it} \\
Nicola M. Pugno\\ \email{nicola.pugno@unitn.it} \\
Jos\'e Roberto de F. Arruda \\ \email{arruda@fem.unicamp.br}
}


\begin{abstract}
The design of structures that can yield efficient sound insulation performance is a recurring topic in the acoustic engineering field. Special attention is given to panels, which can be designed using several approaches to achieve considerable sound attenuation. In a previous work, we have presented the concept of thickness-varying periodic plates with optimised profiles to inhibit flexural wave energy propagation. In this work, motivated by biological structures that present multiple locally-resonant elements able to cause acoustic cloaking, we extend our shape optimisation approach to design panels that achieve improved acoustic insulation performance using either thickness-varying profiles or locally-resonant attachments. The optimisation is performed using numerical models that combine the Kirchhoff plate theory and the plane wave expansion method. Our results indicate that panels based on locally resonant mechanisms have the advantage of being robust against variation in the incidence angle of acoustic excitation and, therefore, are preferred for single-leaf applications.
\end{abstract}

\maketitle

\section{Introduction}

Sound insulation  has been one of the most studied subjects in various fields of Engineering \cite{barron2002industrial,cremer2013korperschall,ginn1978architectural}. Common solutions to achieve this objective include (i) the use of absorptive materials chosen to accept acoustic energy and dissipate it in the form of heat \cite{doutres2005characterisation,doutres2007porous}, and (ii) the use of systems with a large impedance mismatch in the acoustic transmission path, thus reflecting the sound energy, instead of transmitting it \cite{craik2000sound,tadeu2004sound}.

Recent findings in the field of Biology have shown that the wings of moths and butterflies are endowed with specialized double-layered scales with nanostructures that are able to effectively perform ultrasound absorption and create an acoustic camouflage against predators \cite{shen2018biomechanics,clare2015acoustic}. In this case, multiple scales which individually contribute as sub-wavelength resonating elements are connected via a thin membrane to absorb impinging waves over a wide frequency range \cite{neil2020moth}. Such intriguing capabilities may certainly entice the design of novel acoustic metamaterials (MMs) for various engineering applications \cite{yang2017sound}. Especially, a remarkable resemblance with panels with embedded resonators and their application in noise insulation is promptly noticed.

Panels are a common solution for sound insulation in many engineering vibroacoustic applications \cite{long2005architectural}. In this case, the control of noise in the low-frequency regions depends on the stiffness and/or mass of the panel \cite{fahy2007sound}, which may lead to increasingly large panel thickness to improve acoustic performance. To avoid this issue, a number of panel designs such as sandwich \cite{wang2005assessment}, honeycomb \cite{sui2015lightweight}, and fibre-reinforced composite panels \cite{zhang2017sound} are capable of combining high stiffness with low mass. These solutions may further benefit from the use of (i) phononic crystals (PCs) and (ii) periodic MMs, since these may possess phononic band gaps, i.e., frequency ranges where no free wave propagation is allowed in the solid medium \cite{romero2019fundamentals}. Such frequency bands arise typically due to the mechanisms of (i) Bragg scattering \cite{kushwaha1994theory}, in which case the frequency range is associated with the periodicity of the medium, and (ii) local resonance \cite{liu2000locally}, where Fano-like interference is capable of opening band gaps in the sub-wavelength scale \cite{goffaux2002evidence}.

Claeys et al. have discussed the acoustic radiation efficiency of local resonance-based MMs \cite{claeys2014acoustic} and proposed its use in the design of a lightweight acoustic insulator \cite{claeys2016lightweight}. The use of distributed resonators has also been investigated as an option to create MMs for both thin \cite{xiao2012sound} and thick \cite{oudich2014general} plate structures using the plane wave expansion (PWE) method to achieve an improved sound transmission loss (STL). Furthermore, Van Belle et al. have demonstrated \cite{van2019acoustic} that both PCs and MMs are able to improve the STL of infinite plates. For MMs, this reduction occurs inside the band gap regions, due to sub-wavelength vibration suppression, while, for PCs, these occur outside the band gap regions due to specific vibration patterns. However, investigations on the use of corrugated profiles \cite{sorokin2016effects,pelat2019control} to design both PCs and MMs for vibroacoustic applications remain largely unexplored.

In a previous work, we have proposed the optimisation of Fourier coefficients describing the shape of periodic plates for maximising Bragg-type band gaps for structural applications \cite{poggetto2020widening}. In this work, motivated by the characteristics obtained by the combination of resonating elements connected via a thin membrane present in moth wings, we propose the extension of our previously presented optimisation approach to design (i) thickness-varying panels or (ii) panels embedded with multiple resonating elements for sound insulation applications. The optimisation is performed for the normal incidence of impinging waves and is also assessed for the cases of oblique and diffuse incidence. Both single- and double-leaf panels are investigated.

In Section \ref{periodic_fourier}, we revise the concepts relative to thickness-varying plates and periodically distributed resonators and their corresponding Fourier series representation. Section \ref{vibroacoustic_analysis} presents some key concepts and definitions, which are needed for the analysis of the vibroacoustic behaviour. Section \ref{sound_insulation} briefly reviews the formulation for the vibroacoustic behaviour of a single-leaf infinite panel and its extension to the double-leaf case using the PWE method, describing also the metrics related with the calculation of the STL and radiated acoustic pressure, as well as analytical solutions. Section \ref{optimization} presents the optimisation problem, stating its cost function, optimised variables, and constraints, and Section \ref{results} presents the obtained numerical results. Concluding remarks are drawn in Section \ref{conclusion}.

\section{Periodic media and Fourier series representation} \label{periodic_fourier}

\subsection{Plate displacement field}

Even though the sound transmission characteristics of simple panels are commonly obtained considering analytical solutions \cite{kinsler1999fundamentals}, numerical formulations are generally required for complex structures \cite{kang1996finite,yang2017prediction}. The PWE method can be applied to determine the wave propagation characteristics of one-, two- \cite{poggetto2021flexural}, and three-dimensional periodic MMs \cite{poggetto2020elastic}, and its improved computational efficiency, when compared with finite element-based methods \cite{beli2018wave}, further motivates its use for optimisation problems \cite{li2009genetic,doosje2000photonic,bin2011improved}. The PWE method requires the analytical description of the displacement field of the medium and its material/geometric properties using their corresponding Fourier series, which are described in this section.

Consider a plate lying in the $xy$-plane with coordinates described by the two-dimensional position vector $\mathbf{r} = x \hat{\mathbf{i}} + y \hat{\mathbf{j}}$. The plate transverse displacement $u_z(\mathbf{r},t)$ can be described using \cite{poggetto2020widening}
\begin{equation} \label{displacements_fourier}
    u_z(\mathbf{r},t) = e^{-\text{i} \omega t} \sum_{\mathbf{G}} \hat{u}_z(\mathbf{G}) e^{\text{i} (\mathbf{k} +
    \mathbf{G}) \cdot \mathbf{r} } \, ,
\end{equation}
where $\hat{u}_z(\mathbf{G})$ represents spatial Fourier coefficients, the wave propagation in the plate is described by the direction and wavelength given by the wave vector $\mathbf{k} = k_x \hat{\mathbf{i}} + k_y \hat{\mathbf{j}}$, and $\mathbf{G}$ is the reciprocal lattice vector, which is given, for a square plate of side length $L$, by
\begin{equation} \label{G}
 \mathbf{G} = n_x \frac{2\pi}{L} \hat{\mathbf{i}} + n_y \frac{2\pi}{L} \hat{\mathbf{j}} = \mathbf{G}_{n_x, \, n_y} \, ,
\end{equation}
for integers $\{ n_x, \, n_y \} \in \mathbb{Z}$. If indexes $n_x$ and $n_y$ are in the range $[-N_{\max}, N_{\max}]$, one obtains a number of plane waves equal to $n_G = (1+2N_{\max})^2$.

In the case of a thin plate where the effects of rotational inertia and shear strain are negligible, the Kirchhoff plate theory can be used to obtain the corresponding elastodynamic equations, which can be written in the rectangular coordinate system as \cite{ventsel2001thin,leissa1969vibration}
\begin{equation} \label{plate_kirchhoff}
    \small
    \frac{\partial^2 }{\partial x^2} \bigg[
    D \bigg( \frac{\partial^2 u_z}{\partial x^2} + \nu \frac{\partial^2 u_z}{\partial y^2} \bigg) \bigg]  + 
    2 \frac{\partial^2}{\partial x \partial y} \bigg[ D(1-\nu) \frac{\partial^2 u_z}{\partial x \partial y} \bigg] +
    \frac{\partial^2}{\partial y^2} \bigg[ D \bigg( \nu \frac{\partial^2 u_z}{\partial x^2} + \frac{\partial^2 u_z}{\partial y^2} \bigg) \bigg] + \rho h \frac{\partial^2 u_z}{\partial t^2}
    = q(\mathbf{r},t)  \, ,
\end{equation}
where $\rho$ is the material density, $\nu$ is the Poisson's ratio, $h$ is the plate thickness, $D = Eh^3/12(1-\nu^2)$ is the plate flexural stiffness, where $E$ is the Young's modulus, and $q(\mathbf{r},t)$ is the distributed loading on the plate surface, which can include the presence of both fluid loading and the interaction with mass-spring resonators.

The flexural wave speed in plates is given by
\begin{equation} \label{flexural_wave_speed}
 c = \sqrt[4]{ \frac{D}{\rho h} }\sqrt{\omega} \, ,
\end{equation}
which leads to the relation between the wavelength $\lambda$ and frequency $f$, after substituting $c = \lambda f$ and $\omega = 2 \pi f$, written as $\lambda = \sqrt[4]{ \frac{D}{\rho h} } \, \sqrt{ \frac{2\pi}{f} }$, where the relation $h \ll \lambda$ must hold true for all the analysed frequencies for the Kirchhoff plate theory to be valid.

\subsection{Plate thickness}

\begin{figure}[!h]
    \centering
    \includegraphics[scale=0.8]{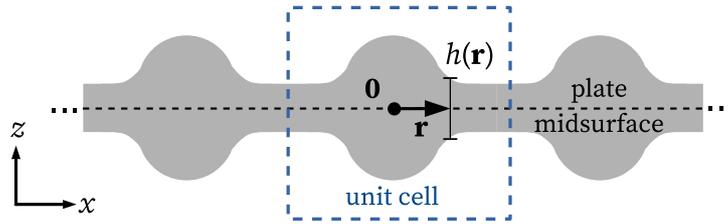}
    \caption{Thickness $h(\mathbf{r})$ for a periodic thin plate lying on the $xy$ plane.}
    \label{plate_thickness_simple}
\end{figure}

The periodic thickness of the thin plate (figure \ref{plate_thickness_simple}) can be described by the spatial-dependent function $h(\mathbf{r})$ given by
\begin{equation} \label{h_fourier}
    h(\mathbf{r}) = \sum_{\mathbf{G}} \hat{h}(\mathbf{G}) \, e^{\text{i} \mathbf{G} \cdot \mathbf{r}} \, ,
\end{equation}
where the Fourier coefficients of $h(\mathbf{r})$ can be written, for integers $n_x$ and $n_y$ (Eq. (\ref{G})) as
\begin{equation} \label{h_fourier2}
 \hat{h}(\mathbf{G}_{n_x, \, n_y}) = \hat{h}_{n_x, \, n_y} \, ,
\end{equation}
leading to the expression of the thickness profile of a plate in terms of the coefficients of the corresponding reciprocal lattices as
\begin{equation} \label{h_fourier3}
 h(\mathbf{r}) = \sum_{n_y} \sum_{n_x} \hat{h}_{n_x, \, n_y} 
 e^{\text{i} \mathbf{G}_{n_x, \, n_y} \cdot \mathbf{r}} \, .
\end{equation}

The space-dependent flexural rigidity of the plate is given by $D(\mathbf{r}) = \frac{E h^3(\mathbf{r})}{12(1-\nu^2)}$, which has the same period as $h(\mathbf{r})$, and can also be expressed in terms of its Fourier series using
\begin{equation} \label{D_fourier}
 D(\mathbf{r}) = \sum_{\mathbf{G}} \hat{D}(\mathbf{G}) \, e^{\text{i} \mathbf{G} \cdot \mathbf{r}} \, ,
\end{equation}
with Fourier coefficients $\hat{D}(\mathbf{G})$ that can be obtained from the coefficients $\hat{h}(\mathbf{G})$ using the procedure described in \cite{poggetto2020widening}.

The set of coefficients $\hat{h}_{n_x, \, n_y}$ can be determined such that
a thickness profile which presents the best possible sound insulation performance is obtained, while respecting imposed geometric constraints. The optimisation problem with its associated metrics and constraints are defined in Section \ref{optimization}.

\subsection{Mass-spring resonators}

\begin{figure}[!h]
    \centering
    \makebox[\textwidth]{
    \includegraphics[scale=0.8]{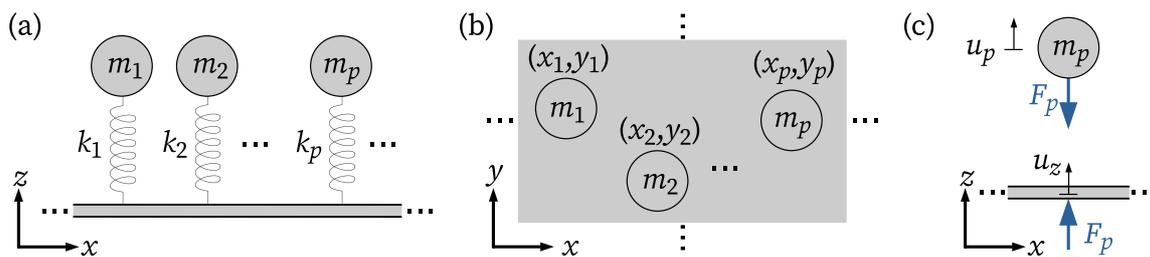}
    }
    \caption{Distribution of mass-spring resonators on a thin plate.
    (a) The $p$-th resonator is composed of a mass $m_p$ and a spring $k_p$
    (b) with coordinates $(x_p, \, y_p)$.
    (c) The displaced mass exerts an out-of-plane force with intensity $F_p$ at the plate.}
    \label{resonators}
\end{figure}

The inclusion of periodic resonator-type structures is an interesting option to control vibrations in the sub-wavelength scale \cite{torrent2013elastic,miranda2019flexural,xiao2012flexural,haslinger2017controlling}. Here we consider a set of $n_r$ independent ideal mass-spring resonators, where each resonator is defined by a point mass $m_p$ and a spring stiffness $k_p$ (figure \ref{resonators}a), fixed to the plate at coordinates $\mathbf{r}_p = x_p \hat{\mathbf{i}} + y_p \hat{\mathbf{j}}$ (figure \ref{resonators}b). Since the plate is periodic, the $p$-th resonator has coordinates which are restricted to the unit cell domain, i.e., $-L/2 < \{ x_p, \, y_p \} < L/2$, with a corresponding out-of-plane displacement, $u_p(\mathbf{r}_p,t)$, exerting a force with intensity $F_p(\mathbf{r}_p,t)$ on the plate (figure \ref{resonators}c), which holds for $p = \{ 1, \, 2, \, \dots, n_r \}$.

The periodic force $q_r(\mathbf{r},t)$ applied by the set of resonators on the plate can be described by
\begin{equation} \label{qr}
 q_r(\mathbf{r},t) = e^{-\text{i} \omega t}
 \sum_p \sum_{\mathbf{R}}
 F_p(\mathbf{r}_p + \mathbf{R}, \, t) \, \delta( \mathbf{r} - (\mathbf{r}_p + \mathbf{R} ) ) \, ,
\end{equation}
where $\mathbf{R} = m_x L \, \hat{\mathbf{i}} + m_y L \, \hat{\mathbf{j}}$, $\{ m_x, \, m_y \} \in \mathbb{Z}$, is the square lattice direct vector, representing the plate periodicity, and $\delta( \mathbf{r} - (\mathbf{r}_p + \mathbf{R} ) ) =
\delta( x - (x_p + m_x L) ) \, \delta( y - (y_p + m_y L ) )$ is the two-dimensional Dirac delta function \cite{xiao2012flexural}, which implies $\delta(\mathbf{r} \neq \mathbf{r}_p + \mathbf{R}) = 0$.

The periodic spring force can be related with the dynamic equation of the $p$-th resonator mass by
\begin{equation} 
 F_p (\mathbf{r}_p + \mathbf{R}, \, t) = k_p ( 
 u_p (\mathbf{r}_p + \mathbf{R} , \, t) - u_z (\mathbf{r}_p + \mathbf{R}, \, t) )
 = - m_p \ddot{u}_p (\mathbf{r}_p + \mathbf{R}, \, t)
 \, ,
\end{equation}
where both the resonator and plate displacements, respectively, $u_p (\mathbf{r},t)$ and $u_z(\mathbf{r},t)$, are evaluated at the points $\mathbf{r}_p + \mathbf{R}$ and can be used, by considering harmonic displacements denoted in the form $f (\mathbf{r}, \, t) = e^{-\text{i}\omega t} f(\mathbf{r})$, to write
\begin{equation} \label{Fp}
 F_p(\mathbf{r},t) = \frac{k_p m_p \omega^2}{k_p - m_p \omega^2} u_z(\mathbf{r},t) \, .
\end{equation}

Equation (\ref{displacements_fourier}) implies
$u_z( \mathbf{r}_p + \mathbf{R} ) = e^{\text{i} \mathbf{k} \cdot \mathbf{R}} u_z( \mathbf{r}_p)$, which can be used with the relation given by \cite{beli2018wave,miranda2019flexural}
\begin{equation} \label{delta_reciprocal}
 \sum_{\mathbf{R}} e^{\text{i} \mathbf{k} \cdot \mathbf{R}}
 \delta( \mathbf{r} - (\mathbf{r}_p + \mathbf{R} ) )
 = \frac{1}{S}
 \sum_{\mathbf{G}} 
 e^{\text{i} (\mathbf{k}+\mathbf{G}) \cdot \mathbf{r}}
 e^{-\text{i} (\mathbf{k}+\mathbf{G}) \cdot \mathbf{r}_p} \, ,
\end{equation}
where $S = L^2$, and combined with Eq. (\ref{qr}) and (\ref{Fp}) using distinct summation indexes, to write
\begin{equation} \label{qr_fourier}
 q_r(\mathbf{r},t) = e^{-\text{i} \omega t}
 \sum_p \frac{k_p m_p \omega^2}{k_p - m_p \omega^2}
 \frac{1}{S}
 \sum_{\mathbf{G}} 
 e^{\text{i} (\mathbf{k}+\mathbf{G}) \cdot \mathbf{r}}
 e^{-\text{i} (\mathbf{k}+\mathbf{G}) \cdot \mathbf{r}_p}
 \,
 \sum_{\mathbf{H}} \hat{u}_z(\mathbf{H}) e^{\text{i} (\mathbf{k} +
    \mathbf{H}) \cdot \mathbf{r}_p } \, .
\end{equation}

This equation can be included as corresponding loading terms to account for locally resonant mechanisms present in the plate.


\section{Fluid-structure interaction} \label{vibroacoustic_analysis}

In this section, we introduce the basic definitions associated with the analysis of the vibroacoustic problem, including the representation of acoustic waves and the fluid-structure coupling formulation. Since we are concerned with acoustic insulation applications, the fluid is generally considered as air, although the presented denomination has been chosen for the sake of generality \cite{yang2017prediction}.

\begin{figure}[!h]
    \centering
    \makebox[\textwidth]{
    \includegraphics[scale=0.8]{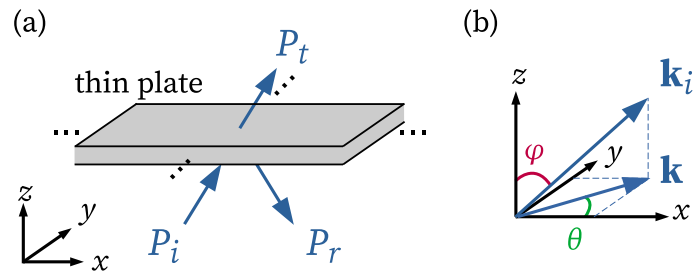}
    }
    \caption{Acoustic plane waves: (a) incident ($P_i$), reflected ($P_r$), and transmitted ($P_t$) waves; (b) wave vector of the incident wave $\mathbf{k}_i$, its component $\mathbf{k}$ in the $xy$-plane, and angles $\varphi$ and $\theta$ for spherical coordinates projection.}
    \label{acoustic_components}
\end{figure}

Consider an incident plane wave ($P_i$) propagating through a fluid, which impinges on an infinite thin plate. The plate immersed in fluid will be referred to as panel. The resulting motion of the plate excites the surrounding fluid, which causes the formation of both reflected ($P_r$) and transmitted ($P_t$) plane waves, respectively, at the same and the opposing sides of the incident face (figure \ref{acoustic_components}a). The wavenumber components of the incident wave can be described using the angles $\varphi$, with respect to the $z$-axis, and its projection in the $xy$-plane with an angle $\theta$, with respect to the $x$-axis (figure \ref{acoustic_components}b).

The incident plane wave can thus be expressed as \cite{yang2017prediction}
\begin{equation}
    P_i(\mathbf{r},z,t) =
    e^{-\text{i} \omega t} \overline{P}_i e^{\text{i} \mathbf{k} \cdot \mathbf{r}} e^{\text{i} k_z z} \, ,
\end{equation}
where $\overline{P}_i$ represents the incident plane wave complex amplitude, the in-plane components of the incident wave vector $\mathbf{k}_i$ are represented by $\mathbf{k}$ and the out-of-plane components by $k_z$, with wavenumber components described in the Cartesian coordinate system using
\begin{equation}
        k_x = k_0 \sin \varphi \cos \theta \, , 
        k_y = k_0 \sin \varphi \sin \theta \, ,
        k_z = k_0 \cos \varphi \, ,
\end{equation}
where $k_0 = |\mathbf{k}_i| = \omega/c_0$ is the wavenumber of the incident wave, with $\omega$ the angular frequency of the incident wave, and $c_0$ the speed of sound in air. The wavelength induced in the plate by the incident plane wave is given by $\sqrt{k_x^2 + k_y^2} = |\mathbf{k}| = k_0 \sin \varphi$ and is named trace wavenumber  \cite{fahy2007sound}, which becomes zero for the case of normal incidence.

The reflected and transmitted waves can be described, respectively, similarly to Eq. (\ref{displacements_fourier}), as \cite{oudich2014general}
\begin{subequations} \label{Pr_Pt}
    \begin{align}
    P_r(\mathbf{r},z,t) &= e^{-\text{i} \omega t} \sum_{\mathbf{G}} \hat{P}_r(\mathbf{G}) e^{\text{i} ( \mathbf{k}+\mathbf{G}) \cdot \mathbf{r}} e^{-\text{i} k_{za}(\mathbf{G}) z} \, , \label{Pr} \\
    P_t(\mathbf{r},z,t) &= e^{-\text{i} \omega t} \sum_{\mathbf{G}} \hat{P}_t(\mathbf{G}) e^{\text{i} ( \mathbf{k}+\mathbf{G}) \cdot \mathbf{r}} e^{\text{i} k_{za}(\mathbf{G}) z} \, , \label{Pt}
    \end{align}
\end{subequations}
where $k_{za}(\mathbf{G})$ is the $z$-direction component of the wave vectors associated with the reflected and transmitted waves, which depend on the wavenumber of the incident wave and the trace wavenumber and can be calculated according to
\begin{subequations}
    \begin{align}
    k_{za}(\mathbf{G}) &= \sqrt{k_0^2 - |\mathbf{k}+\mathbf{G}|^2}\, , \text{ if } k_0^2 \geq |\mathbf{k}+\mathbf{G}|^2 \, , \\
    k_{za}(\mathbf{G}) &= \text{i} \sqrt{|\mathbf{k}+\mathbf{G}|^2 - k_0^2}\, , \text{if } k_0^2 < |\mathbf{k}+\mathbf{G}|^2 \, .
    \end{align}
\end{subequations}

Both reflected and transmitted sound pressures have $n_G$ Fourier coefficients ($\hat{P}_r(\mathbf{G})$ and $\hat{P}_t(\mathbf{G})$, respectively), which must be determined with the application of the fluid-structure coupling equations.

The incident plane wave can be rewritten, keeping the same notation as Eq. (\ref{Pr_Pt}), as
\begin{equation} \label{Pi}
    P_i(\mathbf{r},z,t) =
    e^{-\text{i} \omega t} \sum_{\mathbf{G}} \hat{P}_i(\mathbf{G}) 
    e^{\text{i} (\mathbf{k}+\mathbf{G}) \cdot \mathbf{r}} e^{\text{i} k_z z} \, ,
\end{equation}
where $\hat{P}_i(\mathbf{G}=\mathbf{0}) = \overline{P}_i$ and $\hat{P}_i(\mathbf{G}\neq \mathbf{0}) = 0$.

Henceforth, we shall suppose that the plate is thin enough so that the effects of waves applying pressure at the lower and upper sides of the panel can be approximated by their application at the plate midsurface ($z=0$). Acoustic waves apply pressures in opposing directions (see figure \ref{acoustic_components}a): the incident and reflected waves create a force in the positive ($+z$) direction, while the transmitted wave creates a force in the negative ($-z$) direction. Such fluid loading $q_f = q_f(\mathbf{r},t)$ can be written using Eqs. (\ref{Pr_Pt}) and (\ref{Pi}) as
\begin{equation} \label{qf}
    q_f(\mathbf{r},t) = P_i(\mathbf{r},z=0,t) + P_r(\mathbf{r},z=0,t) - P_t(\mathbf{r},z=0,t) \\
    = e^{-\text{i} \omega t} \sum_{\mathbf{G}}
    \bigg[
    \hat{P}_i (\mathbf{G}) + \hat{P}_r(\mathbf{G}) - \hat{P}_t(\mathbf{G}) 
    \bigg]
    e^{\text{i} ( \mathbf{k}+\mathbf{G}) \cdot \mathbf{r}} \, .
\end{equation}

The continuity of accelerations at the fluid-structure interface can be described as
\begin{equation} \label{acc_continuity}
    \frac{\partial P}{\partial z} \bigg|_{z=z_0} = \rho_0 \omega^2 u_z \big|_{z=z_0} \, ,
\end{equation}
where $z_0$ is the $z$-coordinate of the fluid-structure interface, and $\rho_0$ is the mass density of the fluid. Since here a two-dimensional plate model is used, the transverse displacement is independent of the $z$-coordinate, which implies $ u_z|_{z=z_0} = u_z$.

Equation (\ref{acc_continuity}) can be evaluated for the incident and reflected waves and expanded using Eqs. (\ref{displacements_fourier}), (\ref{Pr}), and (\ref{Pi}), leading to the relation, for each $\mathbf{G}$, given by
\begin{equation} \label{coupling1b}
 \hat{P}_r(\mathbf{G})
  =
  \frac{k_z}{k_{za}(\mathbf{G})} \hat{P}_i(\mathbf{G})
  + \frac{\text{i} \rho_0 \omega^2}{k_{za}(\mathbf{G})} \hat{u}_z(\mathbf{G})
  \, .
\end{equation}

An analogous relation can be used for the coupling at the transmitting face, where the continuity of accelerations and Eqs. (\ref{Pt}) and (\ref{displacements_fourier}) allow to write
\begin{equation} \label{coupling2b}
    \hat{P}_t(\mathbf{G}) = - \frac{\text{i} \rho_0 \omega^2}{k_{za}(\mathbf{G})}  \hat{u}_z(\mathbf{G}) \, .
\end{equation}

Equations (\ref{coupling1b}) and (\ref{coupling2b}) can be substituted in Eq. (\ref{qf}) to write
\begin{equation} \label{qf_final}
    q_f(\mathbf{r},t) = 
    e^{-\text{i} \omega t} \sum_{\mathbf{G}}
    \bigg[
    \bigg( 1 + \frac{k_z}{k_{za}(\mathbf{G})} \bigg) \hat{P}_i (\mathbf{G})
    + 2 \frac{\text{i} \rho_0 \omega^2}{k_{za}(\mathbf{G})} \hat{u}_z(\mathbf{G})
    \bigg]
    e^{\text{i} ( \mathbf{k}+\mathbf{G}) \cdot \mathbf{r}} \, ,
\end{equation}
which represents the equivalent loading imposed on the plate by the incident pressure wave and the fluid-structure interaction.

In the next sections, for the sake of simplicity, the effects of the thickness variation and the inclusion of resonators are presented separately.


\section{Sound insulation panels} \label{sound_insulation}

\subsection{Single-leaf formulation}

Considering the case of a PC, i.e., a thickness-varying plate without the presence of local resonators, Eqs. (\ref{h_fourier}) and (\ref{D_fourier}) can be used with a summation index of $\mathbf{H}$ for material properties, and Eqs. (\ref{qf_final}) and (\ref{displacements_fourier}) with a summation index $\mathbf{G}$ for displacement and acoustic waves in Eq. (\ref{plate_kirchhoff}), one obtains
\begin{equation}
    \begin{aligned}
    & e^{-\text{i} \omega t} \sum_{\mathbf{H}}  \sum_{\mathbf{G}} \bigg[ 
    \hat{D}(\mathbf{H}) \{ [(k_x+G_x)(k_x+G_x+H_x) + (k_y+G_y)(k_y+G_y+H_y)]^2 \\
    & + \nu [ (k_x+G_x)(k_y+G_y+H_y) - (k_y+G_y)(k_x+G_x+H_x) ]^2 \} \hat{u}_z(\mathbf{G}) - \omega^2 \rho \hat{h}(\mathbf{H}) \hat{u}_z(\mathbf{G}) \\
    & 
    - \bigg( 1 + \frac{k_z}{k_{za}(\mathbf{G})} \bigg) \hat{P}_i (\mathbf{G}) \delta(\mathbf{H})
    - 2 \frac{\text{i} \rho_0 \omega^2}{k_{za}(\mathbf{G})} \hat{u}_z(\mathbf{G}) \delta(\mathbf{H})
    \bigg]
    e^{\text{i} (\mathbf{k}+\mathbf{G}+\mathbf{H}) \cdot \mathbf{r}} = 0 \, .
\end{aligned}
\end{equation}

The orthogonality property of the complex exponential \cite{strang1988linear,hsu1967fourier} can be used to write, for every $\mathbf{r}$ and $\mathbf{H}$, a set of linear equations, for $i=1, \, \dots, \, n_G$, obtained by substituting $\mathbf{H} = \mathbf{G}_i - \mathbf{G}_j$ and $\mathbf{G} = \mathbf{G}_j$, written as
\begin{equation} \label{plate_pressures}
    \begin{aligned}
    & \sum_{\mathbf{G}_j} \bigg[ \hat{D}_{ij} \{ [(k_x+G_{xj})(k_x+G_{xi}) + (k_y+G_{yj})(k_y+G_{yi})]^2 \\
    & + \nu [ (k_x+G_{xj})(k_y+G_{yi}) - (k_y+G_{yj})(k_x+G_{xi}) ]^2 \} - \omega^2 \rho \hat{h}_{ij} \bigg] \hat{u}_z(\mathbf{G}_j) \\
    & - \bigg( 1 + \frac{k_z}{k_{za}(\mathbf{G}_i)} \bigg) \hat{P}_i (\mathbf{G}_i) 
    - 2 \frac{\text{i} \rho_0 \omega^2}{k_{za}(\mathbf{G}_i)} \hat{u}_z(\mathbf{G}_i)
    = 0 \, ,
    \end{aligned}
\end{equation}
where $\hat{D}_{ij} = \hat{D}(\mathbf{G}_i - \mathbf{G}_j)$, $\hat{h}_{ij} = \hat{h}(\mathbf{G}_i - \mathbf{G}_j)$. This previous equation can be simplified as
\begin{equation} \label{plate_pressures_u2}
    \sum_{\mathbf{G}_j} \bigg[ \hat{D}_{ij} f_{ij} - \omega^2 \rho \hat{h}_{ij} \bigg] \hat{u}_z(\mathbf{G}_j)
    - 2 \frac{\text{i} \rho_0 \omega^2}{k_{za}(\mathbf{G}_i)}
    \hat{u}_z(\mathbf{G}_i)
    = \bigg( 1 + \frac{k_z}{k_{za}(\mathbf{G}_i)} \bigg)
    \hat{P}_i(\mathbf{G}_i) \, ,
\end{equation}
where $f_{ij} = f_{ij}(\mathbf{k},\mathbf{G}_i,\mathbf{G}_j)$ is given by
\begin{equation} \label{f_ij}
 \small
 f_{ij} = [(k_x+G_{xj})(k_x+G_{xi}) + (k_y+G_{yj})(k_y+G_{yi})]^2 + \nu [ (k_x+G_{xj})(k_y+G_{yi}) - (k_y+G_{yj})(k_x+G_{xi}) ]^2 \, .
\end{equation}

Equation (\ref{plate_pressures}) represents a set of $n_G$ linear equations, which can be rewritten in matrix form  last relation allows to determine the Fourier components $\hat{u}_z(\mathbf{G})$ using
\begin{equation} \label{set_u_wavy}
 (\widetilde{\mathbf{D}} + \mathbf{D}_f ) \hat{\mathbf{u}}_z = \mathbf{f} \, ,
\end{equation}
where the matrix $\widetilde{\mathbf{D}}$ can be written as
\begin{equation} \label{set_u_wavy_D}
 \widetilde{\mathbf{D}} = \widetilde{\mathbf{K}} - \omega^2 \widetilde{\mathbf{M}} \, ,
\end{equation}
with the components of matrices $\widetilde{\mathbf{K}}$, $\widetilde{\mathbf{M}}$, the diagonal matrix $\mathbf{D}_f$, and the vector $\mathbf{f}$ are respectively given by
\begin{equation} \label{set_transmission_wavy}
 (\widetilde{\mathbf{K}})_{ij} = \hat{D}_{ij} f_{ij} \, , 
 (\widetilde{\mathbf{M}})_{ij} = \rho \hat{h}_{ij} \, , 
 (\mathbf{D}_f)_{ii} = - 2 \frac{\text{i} \rho_0 \omega^2}{ k_{za}(\mathbf{G}_i)} \, , 
 (\mathbf{f})_i = \bigg( 1 + \frac{k_z}{k_{za}(\mathbf{G}_i)} \bigg) \hat{P}_i(\mathbf{G}_i) \, , 
\end{equation}
and the vector of unknowns is given by
\begin{equation} \label{vector_unkowns}
 \hat{\mathbf{u}}_z = \left\{ \hat{u}_z(\mathbf{G}_1), \, \dots, \, \hat{u}_z(\mathbf{G}_{n_G}) \right\}^T \, .
\end{equation}

In Eq. (\ref{set_u_wavy}), matrix $\widetilde{\mathbf{D}}$ accounts for the periodic plate dynamic stiffness characteristics, matrix $\mathbf{D}_f$ includes the additional impedance associated with the fluid-structure interaction, and vector $\mathbf{f}$ represents the excitation induced by the incident wave pressure. Finally, the vector of Fourier coefficients of the transmitted wave can be represented as
\begin{equation}
 \hat{\mathbf{P}}_t = \left\{ \hat{P}_t(\mathbf{G}_1), \, \dots, \, \hat{P}_t(\mathbf{G}_{n_G}) \right\}^T \, ,
\end{equation}
which can be directly obtained from the solution of Eq. (\ref{set_u_wavy}) with the use of Eq. (\ref{coupling2b}). Thus, for each pair of angles indicating the direction of the incident wave ($\varphi$, $\theta$), Eq. (\ref{set_u_wavy}) can be solved for a given set of angular frequencies, $\omega$, thus yielding the Fourier components relative to the plate displacements, $u_z(\mathbf{r},t)$, and, consequently, transmitted waves, $P_t(\mathbf{r},z,t)$. This formulation is similar to that presented by Xiao et al. \cite{xiao2012sound}, with the proper use of Fourier series coefficients corresponding to the plate thickness variation.

The dispersion characteristics of the wave propagation in the plate can be obtained by neglecting the terms associated with fluid loading and external excitation in Eq. (\ref{set_u_wavy}) (i.e., $\mathbf{D}_f=\mathbf{0}$ and $\mathbf{f} = \mathbf{0}$, respectively), thus obtaining the eigenproblem stated as
\begin{equation}
 \widetilde{\mathbf{K}} \hat{\mathbf{u}}_z = \omega^2 \widetilde{\mathbf{M}} \hat{\mathbf{u}}_z \, ,
\end{equation}
which can be solved for $\omega = \omega(\mathbf{k})$ scanning the contour of the irreducible Brillouin zone (i.e., restricting the wave vector to the contour of the region defined by the high-symmetry points $\Gamma$ $(0, \, 0)$, X $(\pi/L, \, 0)$, and M $(\pi/L, \, \pi/L)$ in the reciprocal space) for a single unit cell of the thickness-varying plate periodic structure (see \cite{poggetto2020widening} for details).

Now considering the case of an elastic MM, i.e., a constant-thickness plate with resonators under fluid loading, Eq. (\ref{plate_kirchhoff}) can be used for the case of constant thickness and the presence of mass-spring resonators, obtaining
\begin{equation} \label{flat_plate_dynamics}
    D \bigg( \frac{\partial^4 u_z}{\partial x^4}
    + 2 \frac{\partial^4 u_z}{\partial^2 x \partial^2 y}
    + \frac{\partial^4 u_z}{\partial y^4} \bigg)
    + \rho h \frac{\partial^2 u_z}{\partial t^2}
    = q_f(\mathbf{r},t) + q_r(\mathbf{r},t) \, ,
\end{equation}
which can be combined with Eqs. (\ref{displacements_fourier}), (\ref{qr_fourier}), and (\ref{qf_final}) to write
\begin{equation}
\footnotesize
\mathclap{
\begin{aligned}
 &\sum_{\mathbf{G}} \bigg[
 D [ (k_x + G_x)^2 + (k_y + G_y)^2 ]^2 - \omega^2 \rho h \bigg] \hat{u}_z(\mathbf{G})
 e^{\text{i} ( \mathbf{k}+\mathbf{G}) \cdot \mathbf{r}}
 =
  \sum_{\mathbf{G}}
    \bigg[
    \bigg( 1 + \frac{k_z}{k_{za}(\mathbf{G})} \bigg) \hat{P}_i (\mathbf{G})
    + 2 \frac{\text{i} \rho_0 \omega^2}{k_{za}(\mathbf{G})} \hat{u}_z(\mathbf{G})
    \bigg]
    e^{\text{i} ( \mathbf{k}+\mathbf{G}) \cdot \mathbf{r}} +
 \\
 &\sum_p \frac{k_p m_p \omega^2}{k_p - m_p \omega^2}
 \frac{1}{S}
 \sum_{\mathbf{G}} 
 e^{\text{i} (\mathbf{k}+\mathbf{G}) \cdot \mathbf{r}}
 e^{-\text{i} (\mathbf{k}+\mathbf{G}) \cdot \mathbf{r}_p}
 \,
 \sum_{\mathbf{H}} \hat{u}_z(\mathbf{H}) e^{\text{i} (\mathbf{k} +
    \mathbf{H}) \cdot \mathbf{r}_p } \, .
\end{aligned}
}
\end{equation}

The orthogonality property of the complex exponentials can once again be used to write an equation analogous to Eq. (\ref{plate_pressures}), leading to
\begin{equation}
\small
\begin{aligned}
 &\bigg[
 D [ (k_x + G_{xi})^2 + (k_y + G_{yi})^2 ]^2 - \omega^2 \rho h
 - 2 \frac{\text{i} \rho_0 \omega^2}{k_{za}(\mathbf{G}_i)}
 \bigg] \hat{u}_z(\mathbf{G}_i)
 \\
 & +
 \sum_{\mathbf{G}_j}
 \bigg[
 \sum_p \frac{k_p m_p \omega^2}{m_p \omega^2 - k_p}
 \frac{1}{S}
 e^{-\text{i} (\mathbf{G}_i - \mathbf{G}_j) \cdot \mathbf{r}_p}
 \bigg]
 \hat{u}_z(\mathbf{G}_j) 
 = \bigg( 1 + \frac{k_z}{k_{za}(\mathbf{G}_i)} \bigg) \hat{P}_i (\mathbf{G}_i) \, .
\end{aligned}
\end{equation}

The previous equation can be rewritten in the matrix form as
\begin{equation} \label{set_u_reso}
 ( \mathbf{D} + \mathbf{D}_r + \mathbf{D}_f ) \hat{\mathbf{u}}_z = \mathbf{f} \, ,
\end{equation}
where the matrix $\mathbf{D}$ can be written as
\begin{equation}
 \mathbf{D} = \mathbf{K} - \omega^2 \mathbf{M} \, ,
\end{equation}
with the diagonal matrices $\mathbf{K}$ and $\mathbf{M}$, and matrix $\mathbf{D}_r$ having their terms respectively given by
\begin{equation} \label{set_transmission_reso}
\begin{aligned}
 (\mathbf{K})_{ii} &= D [ (k_x + G_{xi})^2 + (k_y + G_{yi})^2 ]^2 \, ,
 (\mathbf{M})_{ii} = \rho h \, , \\
 (\mathbf{D}_r)_{ij} &= \sum_p \frac{k_p m_p \omega^2}{m_p \omega^2 - k_p}
 \frac{1}{S}
 e^{-\text{i} (\mathbf{G}_i-\mathbf{G}_j) \cdot \mathbf{r}_p}
 \, .
\end{aligned}
\end{equation}

In Eq. (\ref{set_u_reso}), matrix $\mathbf{D}_r$ represents the dynamic contribution of the distributed resonators, which is superposed to the matrix that represents the constant-thickness plate dynamic stiffness matrix, $\mathbf{D}_r$. It is also interesting to notice, in comparison with Eq. (\ref{set_transmission_wavy}), that matrices $\mathbf{K}$ and $\mathbf{M}$ correspond, respectively, to the diagonals of matrices $\widetilde{\mathbf{K}}$ and $\widetilde{\mathbf{M}}$, while matrix $\mathbf{D}_f$ and vector $\mathbf{f}$ remain the same. A similar problem can be formulated to derive the dispersion relation for a given configuration of resonators, as described in the Supplementary Material.

Equation (\ref{set_u_reso}) accounts for the inclusion of periodically distributed resonators, represented by the term $\mathbf{D}_r$, on a constant-thickness plate whose dynamic stiffness behaviour is represented by the term $\mathbf{D}$. An analogy with Eq. (\ref{set_u_wavy}) suggests that the general case, i.e., with the inclusion of resonators on a plate with a varying thickness profile, can be obtained by performing the substitution $\mathbf{D} \rightarrow \widetilde{\mathbf{D}}$, leading to
\begin{equation} \label{set_u_general}
 ( \widetilde{\mathbf{D}} + \mathbf{D}_r + \mathbf{D}_f ) \hat{\mathbf{u}}_z = \mathbf{f} \, .
\end{equation}

Equation (\ref{set_u_general}) can therefore be used as a general formulation of the STL of the infinite panel, from which the simplified cases of the PC and elastic MM can be derived by reducing to Eq. (\ref{set_u_wavy}) or Eq. (\ref{set_u_reso}), respectively.

\subsection{Extension to double-leaf panels}

\begin{figure}[!h]
    \centering
    \makebox[\textwidth]{
    \includegraphics[scale=0.8]{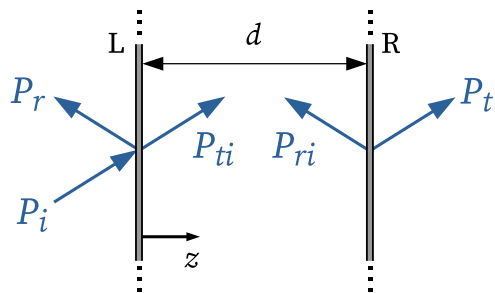}
    }
    \caption{Double-leaf panel with leaves at a distance $d$ away from each other.
    Incident ($P_i$) and reflected ($P_r$) waves are formed before the left leaf (L), while transmitted waves ($P_t$) are formed after the right leaf (R).
    Standing waves ($P_{ti}$ and $P_{ri}$) are developed inside the cavity filled with fluid.}
    \label{double_leaf}
\end{figure}

Consider now the case of a double-leaf sound insulation panel, composed by thin plates labelled as L and R, respectively located at $z=0$ and $z=d$, immersed in a fluid, as depicted in figure \ref{double_leaf}.

In this case, it is necessary to distinguish between the transverse displacements associated with panel leaves L and R, and also discriminate the standing waves $P_{ti}$ and $P_{ri}$, that are developed inside the fluid cavity, described by
\begin{subequations} \label{disp_Pi_Pr_Pt_L}
    \begin{align}
    u_z^{(\text{L})}(\mathbf{r},t) &= e^{-\text{i} \omega t} \sum_{\mathbf{G}} \hat{u}_z^{(\text{L})}(\mathbf{G}) e^{\text{i} (\mathbf{k} +
    \mathbf{G}) \cdot \mathbf{r} } \, , \\
    u_z^{(\text{R})}(\mathbf{r},t) &= e^{-\text{i} \omega t} \sum_{\mathbf{G}} \hat{u}_z^{(\text{R})}(\mathbf{G}) e^{\text{i} (\mathbf{k} +
    \mathbf{G}) \cdot \mathbf{r} } \, , \\
    P_{ti}(\mathbf{r},z,t) &= e^{-\text{i} \omega t} \sum_{\mathbf{G}}
    \hat{P}_{ti}(\mathbf{G})
    e^{\text{i} ( \mathbf{k}+\mathbf{G}) \cdot \mathbf{r}} e^{\text{i} k_{za}(\mathbf{G}) z} \, , \\
    P_{ri}(\mathbf{r},z,t) &= e^{-\text{i} \omega t} \sum_{\mathbf{G}}
    \hat{P}_{ri}(\mathbf{G})
    e^{\text{i} ( \mathbf{k}+\mathbf{G}) \cdot \mathbf{r}} e^{-\text{i} k_{za}(\mathbf{G}) z} \, .
    \end{align}
\end{subequations}

Following the same reasoning as in the previous section, we will begin by writing the fluid-structure coupling equations that allow to express the Fourier coefficients of the plane waves of interest.

Considering the continuity of accelerations (Eq. (\ref{acc_continuity})) at the incident face of panel leaf L and at the transmitted face of panel leaf R, one has, respectively, the relations
\begin{subequations} \label{fluid_structure_LR}
 \begin{align}
    \hat{P}_r(\mathbf{G}) &= 
    \frac{k_{z}}{k_{za}(\mathbf{G})}
    \hat{P}_i(\mathbf{G})
    +
    \frac{\text{i} \rho_0 \omega^2}{k_{za}(\mathbf{G})}
    \hat{u}_z^{(\text{L})}(\mathbf{G})
    \, , \\
    \hat{P}_t(\mathbf{G}) &= -\frac{\text{i} \rho_0 \omega^2}{ k_{za}(\mathbf{G}) e^{\text{i} k_{za}(\mathbf{G}) d}} \hat{u}_z^{(\text{R})}(\mathbf{G}) \, .
 \end{align}
\end{subequations}


The standing waves can be related in an analogous way, considering the continuity of accelerations at the transmitted face of panel leaf L and incident face of panel leaf R, which leads to the the Fourier coefficients of the plane waves inside the acoustic cavity expressed as
\begin{subequations} \label{fluid_structure_internal3}
 \begin{align}
    \hat{P}_{ti}(\mathbf{G}) &=
    \frac{\rho_0 \omega^2}{2 k_{za}(\mathbf{G}) \sin(k_{za}(\mathbf{G})d)}
    \bigg(
    e^{-\text{i} k_{za}(\mathbf{G}) d} \hat{u}_z^{(\text{L})}(\mathbf{G})
    -
    \hat{u}_z^{(\text{R})}(\mathbf{G})
    \bigg)
    \, , \\
    \hat{P}_{ri}(\mathbf{G}) &=
    \frac{\rho_0 \omega^2}{2 k_{za}(\mathbf{G}) \sin(k_{za}(\mathbf{G})d)}
    \bigg(
    e^{\text{i} k_{za}(\mathbf{G}) d} \hat{u}_z^{(\text{L})}(\mathbf{G})
    -
    \hat{u}_z^{(\text{R})}(\mathbf{G})
    \bigg)
    \, .
 \end{align}
\end{subequations}


Equations (\ref{fluid_structure_LR}) and (\ref{fluid_structure_internal3}) can now be coupled with the dynamic equations of both plates. 
The fluid loading on panel leaf L can be written as
\begin{equation}
    q^{(\text{L})}(\mathbf{r},t)
    =
    P_i(\mathbf{r},z=0,t) + P_r(\mathbf{r},z=0,t)
    - P_{ti}(\mathbf{r},z=0,t) - P_{ri}(\mathbf{r},z=0,t) \, ,
\end{equation}
which can be combined with Eqs. (\ref{fluid_structure_LR}) and (\ref{fluid_structure_internal3}) to derive an equation analogous to Eq. (\ref{plate_pressures}) for panel leaf L, which can be stated as
\begin{equation} \label{plate_pressures_L}
    \begin{aligned}
    & \sum_{\mathbf{G}_j} \bigg[ \hat{D}^{(\text{L})}_{ij} f_{ij} - \omega^2 \rho \hat{h}^{(\text{L})}_{ij} \bigg] \hat{u}_z^{(\text{L})}(\mathbf{G}_j)
    + \frac{\rho_0 \omega^2}{ k_{za}(\mathbf{G}_i)}
    \bigg( - \text{i} + \frac{1}{\tan(k_{za}(\mathbf{G}_i)d)} \bigg)
    \hat{u}_z^{(\text{L})}(\mathbf{G}_i)
    \\
    &
    - \frac{\rho_0 \omega^2}{k_{za}(\mathbf{G}_i) \sin(k_{za}(\mathbf{G}_i)d)}
    \hat{u}_{z\mathbf{k}}^{(\text{R})} (\mathbf{G}_i)
    = 
    \bigg( 1 + \frac{k_{z}}{k_{za}(\mathbf{G}_i)} \bigg)
    \hat{P}_i(\mathbf{G}_i)
    \, ,
    \end{aligned}
\end{equation}
where $\hat{D}^{(\text{L})}_{ij}$ and $\hat{h}^{(\text{L})}_{ij}$ refer to the Fourier coefficients of the flexural stiffness and thickness of panel leaf L computed for the reciprocal lattice vectors $\mathbf{G}_i - \mathbf{G}_j$.


Analogously, the fluid loading on panel leaf R can be written as
\begin{equation}
    q^{(\text{R})}(\mathbf{r},t) =
    P_{ti}(\mathbf{r},z=d,t)
    + P_{ri}(\mathbf{r},z=d,t)
    - P_t(\mathbf{r},z=d,t) \, ,
\end{equation}
thus leading to an equation analogous to Eq. (\ref{plate_pressures_L}) for panel leaf R, which reads 
\begin{equation} \label{plate_pressures_R}
    \begin{aligned}
    & \sum_{\mathbf{G}_j} \bigg[ \hat{D}^{(\text{R})}_{ij} f_{ij} - \omega^2 \rho \hat{h}^{(\text{R})}_{ij} \bigg] \hat{u}_z^{(\text{R})}(\mathbf{G}_j)
    - \frac{\rho_0 \omega^2}{k_{za}(\mathbf{G}_i) \sin(k_{za}(\mathbf{G}_i)d)}
    \hat{u}_z^{(\text{L})}(\mathbf{G}_i)
    \\
    &
    + \frac{\rho_0 \omega^2 }{k_{za}(\mathbf{G}_i)}
    \bigg( - \text{i} + \frac{1}{\tan(k_{za}(\mathbf{G}_i)d)} \bigg) \hat{u}_z^{(\text{R})}(\mathbf{G}_i)
    = 0
    \, ,
    \end{aligned}
\end{equation}
where $\hat{D}^{(\text{R})}_{ij}$ and $\hat{h}^{(\text{R})}_{ij}$ have the same meaning as in Eq. (\ref{plate_pressures_L}), but for panel leaf R.

Equations (\ref{plate_pressures_L}) and (\ref{plate_pressures_R}) can be organized in the form of a linear system as
\begin{equation} \label{set_u_double}
 \left[
 \begin{array}{cc}
  \widetilde{\mathbf{D}}^{(L)} + \mathbf{D}_f^{(d)} & \mathbf{D}_f^{(c)} \\
  \mathbf{D}_f^{(c)} & \widetilde{\mathbf{D}}^{(R)} + \mathbf{D}_f^{(d)}
 \end{array}
 \right]
 \left\{
 \begin{array}{c}
  \hat{\mathbf{u}}_z^{(\text{L})} \\
  \hat{\mathbf{u}}_z^{(\text{R})}
 \end{array}
 \right\}
 =
 \left\{
 \begin{array}{cc}
  \mathbf{f} \\
  \mathbf{0}
 \end{array}
 \right\} \, ,
\end{equation}
where matrices $\widetilde{\mathbf{D}}^{(\text{L})} = \widetilde{\mathbf{K}}^{(\text{L})} - \omega^2 \widetilde{\mathbf{M}}^{(\text{L})}$, $\widetilde{\mathbf{D}}^{(\text{R})} = \widetilde{\mathbf{K}}^{(\text{R})} - \omega^2 \widetilde{\mathbf{M}}^{(\text{R})}$, are given by Eq. (\ref{set_transmission_wavy}) for panel leaves L and R, respectively, the loading vector $\mathbf{f}$ is also given by the same equation, and diagonal matrices $\mathbf{D}_f^{(d)}$, $\mathbf{D}_f^{(c)}$ have elements given by
\begin{equation} \label{set_transmission_double}
 \begin{aligned}
  (\mathbf{D}_f^{(d)})_{ii} &= \frac{\rho_0 \omega^2}{ k_{za}(\mathbf{G}_i)}
    \bigg( - \text{i} + \frac{1}{\tan(k_{za}(\mathbf{G}_i)d)} \bigg) \, , \\
  (\mathbf{D}_f^{(c)})_{ii} &= - \frac{\rho_0 \omega^2}{k_{za}(\mathbf{G}_i) \sin(k_{za}(\mathbf{G}_i)d)} \, ,
 \end{aligned}
\end{equation}
and the vectors of unknowns correspond to the Fourier coefficients for the displacements of panel leaves L and R (see Eq. (\ref{vector_unkowns})).

Interestingly, matrix $\mathbf{D}_f^{(d)}$ accounts for the fluid loading impedance at one side ($\mathbf{D}_f$/2, see Eq. (\ref{set_transmission_wavy})) and a contribution for the cavity impedance ($\tan$ term); meanwhile, the coupling between panel leaves L and R is provided by matrix $\mathbf{D}_f^{(c)}$. The form of Eq. (\ref{set_transmission_double}) suggests that it can be easily extended for an arbitrary number of leaves.


\subsection{Acoustic metrics} \label{sound_transmission_loss}

The sound power transmission coefficient ($\tau$) can be calculated for a given set of angles ($\varphi$, $\theta$) and angular frequency $\omega$ using \cite{xiao2012sound}
\begin{equation} \label{tau}
    \tau(\varphi, \, \theta,  \, \omega) = \frac{ \sum_{\mathbf{G}} |\hat{P}_t(\mathbf{G})|^2 \, \text{Re}(k_{za}(\mathbf{G})) }{ |\hat{P}_i|^2 k_z } \, .
\end{equation}

Thus, the sound transmission loss (STL) can be obtained using
\begin{equation} \label{STL}
    \text{STL}(\varphi, \, \theta,  \, \omega) = 10 \log \bigg( \frac{1}{\tau(\varphi, \, \theta,  \, \omega)} \bigg) \, .
\end{equation}

For the case of acoustic waves with oblique incidence angles, the diffuse power transmission coefficient ($\tau_{\text{d}}$) may be calculated using \cite{hambric2016engineering}
\begin{equation} \label{tau_d}
 \tau_{\text{d}}(\theta, \, \omega) = \int_{0}^{\pi/2} \, \tau(\varphi, \, \theta,  \, \omega) \, \sin 2 \varphi \, d\varphi \, ,
\end{equation}
which, for the purpose of numerical evaluation, can be evaluated using angles between $0^o$ and $78^o$ \cite{fahy2007sound}. This definition can also be used with Eq. (\ref{STL}) to express the diffuse STL as
\begin{equation} \label{STL_d}
    \text{STL}_{\text{d}}(\theta, \, \omega) = 10 \log \bigg( \frac{1}{\tau_d(\theta, \, \omega)} \bigg) \, .
\end{equation}

With the purpose of validating the PWE formulation, analytical expressions obtained from the literature are also presented. The analytical sound power transmission coefficient ($\tau_{\text{a}}$) for an infinite constant-thickness flat plate (and thus independent of $\theta$) with the same fluid on both sides can be calculated using \cite{hambric2016engineering}
\begin{equation} \label{tau_a_single}
    \tau_{\text{a}}^{\text{s}}(\varphi, \, \omega) = \frac{(2 \rho_0 c_0 \sec \varphi)^2}{(2 \rho_0 c_0 \sec \varphi)^2 + (\omega \rho h - (D/\omega)(k_0 \sin \varphi)^4 )^2 } \, ,
\end{equation}
which can be used in the place of the numerically obtained $\tau(\varphi, \, \theta,  \, \omega)$ in Eq. (\ref{STL}) to calculate the analytical STL.

The frequency $f_{co}$ given by
\begin{equation} \label{f_co}
    f_{co}(\varphi) = \frac{1}{2\pi} \bigg( \frac{c_0}{\sin \varphi} \bigg)^2
    \sqrt{ \frac{\rho h}{D} } \, ,
\end{equation}
is named coincidence frequency \cite{fahy2007sound}, and corresponds to the frequency at which the structural impedance for the plate is minimal for a given incidence angle ($\varphi$), causing a dip in the STL curve. For $\varphi = \pi/2$ this frequency becomes minimal, and is named critical frequency.

For the case of a double-leaf panel with plates of constant thickness, the power transmission coefficient is given by
\begin{equation} \label{tau_a_double}
    \tau_{\text{a}}^{\text{d}}(\varphi, \, \omega) =
    \bigg|
    \frac{2 \text{i} (\rho_0 c_0 \sec \varphi)^2 \sin(k_0 d \cos \varphi)}{(\text{i} \omega \rho h_{\text{L}} + z_0)(\text{i} \omega \rho h_{\text{R}} + z_0)\sin^2 (k_0 d \cos \varphi) + (\rho_0 c_0 \sec \varphi)^2}
    \bigg|^2
    \, ,
\end{equation}
where $z_0 = \rho_0 c_0 \sec \varphi ( 1 - \text{i} \cot ( k_0 d \cos \varphi))$ is a term associated with the cavity impedance and $h_{\text{L}}$ and $h_{\text{R}}$ are the thicknesses of the first and second leaves, respectively.


\section{Optimisation problem} \label{optimization}

In the low-frequency range, a common solution to improve the STL consists of increasing the panel thickness. However, this is not the most economical solution, since it certainly implies more mass and the use of more material. Thus, the proposed objective here is to determine either an optimal thickness profile distribution or the inclusion of mass-spring resonators to achieve the maximum STL at a given frequency of interest. Thus, this optimisation objective can be stated as 
\begin{equation} \label{optimization_metric}
    \underset{\mathbf{d}}{\text{maximize}}
    \: \phi =
    \int_{\omega_{\min}}^{\omega_{\max}} \,
    \frac{1 - e^{-\Delta\text{STL}(\omega)}}{1 + e^{-\Delta\text{STL}(\omega)}}
     \, d\omega
    \, ,
\end{equation}
where $\Delta\text{STL}(\omega) = \text{STL}(\varphi=0,\omega) - \text{STL}_{\text{ref}}(\varphi=0,\omega)$ is the difference between the normal STL obtained for a set $\mathbf{d}$ of design variables and the reference STL computed for an initial constant-thickness plate, integrated over the frequency range $[ \omega_{\min}, \, \omega_{\max}]$. The set of design variables $\mathbf{d}$ may refer either to a plate thickness profile ($h_{n_x, \, n_y}$, see Eq. (\ref{h_fourier2})) or to a set of mass-spring resonators ($k_p$, $m_p$, $\mathbf{r}_p$, see Eq. (\ref{qr_fourier})). The proposed integrand becomes $1$ for $\Delta\text{STL} \gg 0$ and $-1$ for $\Delta\text{STL} \ll 0$, which indicates an improvement (degradation) with respect to the original STL for sufficiently larger (smaller) values. It is important to notice that the incidence angle $\varphi=0$ does not contribute to the computation of the diffuse STL (Eqs. (\ref{tau_d}) and (\ref{STL_d})), which must be verified to assess the improvement in performance in this case.

The definition of the constraints will depend on the case treated. Therefore, we will describe the thickness-varying plate without resonators case and the constant-thickness plate with resonators case separately hereafter.

\subsection{Thickness-varying plate without resonators} \label{optimization_pc_plate}

Starting from Eq. (\ref{h_fourier3}), the plate thickness may be rewritten explicitly in terms of the $x$ and $y$ coordinates using Eq. (\ref{G}) as
\begin{equation} \label{h_fourier4}
 h(x,y) = \sum_{n_y} \sum_{n_x} \hat{h}_{n_x, \, n_y} 
 e^{\text{i} n_x (2\pi/L)  x} e^{\text{i} n_y (2\pi/L)  y} \, ,
\end{equation}
which presents a considerable reduction in the number of variables $\hat{h}_{n_x, \, n_y}$ needed to describe the plate shape when assuming a symmetry of coefficients centered around $\hat{h}_{0, \, 0}$, i.e., $\hat{h}_{n_x, \, n_y} = \hat{h}_{n_x, \, -n_y} = \hat{h}_{-n_x, \, n_y} = \hat{h}_{-n_x, \, -n_y}$.

The constraints on the plate thickness can be written as
$ h_{\min} \leq h(x,y) \leq h_{\max} $, where $h_{\min}$ and $h_{\max}$ are the minimum and maximum allowed values of plate thickness, respectively, which must hold true for $-L/2 \leq \{ x, \, y \} \leq L/2$. An additional constraint can be imposed to evaluate different plate thickness profiles that present the same mass per unit area. This can be achieved by setting a fixed mean value for the plate thickness, which means material addition implies in the same amount of material removal. Thus, this does not imply in an additional constraint, but in the removal of $\hat{h}_{0, \,0}$ from the list of optimisation variables, fixing it at the beginning of the optimisation process. This leaves us with the suitable boundaries of optimisation variables given by
\begin{equation} \label{hij_constraints}
 -h_{\max} \leq \, \hat{h}_{n_x, \, n_y} \leq h_{\max} \, , \forall \{n_x, \, n_y\} \neq \{0,0\} \, .
\end{equation}

The interested reader is referred to a thorough discussion on this approach \cite{poggetto2020widening}.

\subsection{Constant-thickness plate with resonators}  \label{optimization_mm_plate}

\begin{figure}[!h]
    \centering
    \makebox[\textwidth]{
    \includegraphics[scale=0.8]{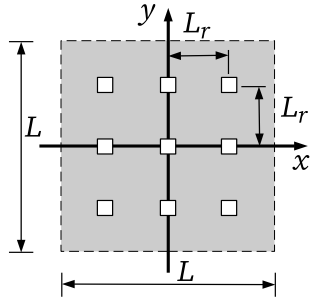}
    }
    \caption{Unit cell with dimensions $L \times L$ with evenly spaced resonators at a distance $L_r$ from each other.}
    \label{resonators_distribution}
\end{figure}

For the case of the inclusion of resonators, some simplifications may be assumed to reduce the number of optimisation variables. First, by recalling that the medium is periodic, let us assume that the resonating structures are equally distributed through the unit cell. Thus, let us assume that the locations of the $p$-th resonator is uniquely determined by
\begin{equation}
 x_p = -L/2 + L_r n_x \, , 
 y_p = -L/2 + L_r n_y \, ,
\end{equation}
where $L_r = L/(n_{rd}+1)$ is the spacing between consecutive resonators in a given direction $x$ or $y$ (figure \ref{resonators_distribution}), where the total number of resonators is given by $n_r = n_{rd}^2$, and $0 \leq \{ n_x , \, n_y \} \leq n_{rd}$.

It is also reasonable to assume that the mass of each resonator is constant, 
i.e., $m_p = \Delta m/n_r$, $\forall p$, where $\Delta m$ is the total added mass of the resonators. To keep the same base of comparison between distinct designs, we also restrict the total amount of added mass to keep the same mass per unit cell. This can be achieved by setting
\begin{equation}
 \Delta m = \rho L^2 \Delta h \, ,
\end{equation}
where $\Delta h = h_0 - h$ is the thickness variation of the plate, calculated as the difference between the current thickness $h$ and the initial thickness $h_0$. Thus, for a plate with constant thickness $h < h_0$, the mass of the resonators can be determined.

The stiffness $k_p$ of each resonator can be set by properly choosing their resonant frequency, $\omega_p$, i.e., $k_p = m_p \omega_p^2$, for $p = \{1 , 2, \, \dots, n_r \}$. The resonant frequency of each resonator can be obtained by sampling a two-dimensional function of a continuous stiffness written as
\begin{equation}
 \omega_p = \omega_c(x_p, \, y_p) \,
\end{equation}
where the continuous function $\omega_c(x,y)$ can be expressed in the same way as Eq. (\ref{h_fourier4}), i.e.,
\begin{equation}
 \omega_c(x,y) = \sum_{n_y} \sum_{n_x}
 \hat{\omega}_{n_x, \, n_y} e^{\text{i} n_x (2\pi/L) x} e^{\text{i} n_x (2\pi/L) y} \, .
\end{equation}

Expressing a continuous resonant frequency function in such a manner may present great advantages for a large number of resonators, since the $\omega_c(x,y)$ function may be sampled at will without increasing the number of design variables. The lower and upper bounds of $\omega_c(x,y)$ may be selected as
\begin{equation}
 \omega^{(r)}_{\min} \leq \omega_c(x,y) \leq \omega^{(r)}_{\max} \, ,
\end{equation}
where $\omega^{(r)}_{\min}$ and $\omega^{(r)}_{\max}$ refer to the minimum and maximum frequencies of interest, respectively. The constraints on the stiffness function can be set in an analogous way as used for thickness, i.e., Eq. (\ref{hij_constraints}), using
\begin{subequations} \label{wij_constraints}
\begin{align}
 \omega^{(r)}_{\min} \leq \, &\hat{\omega}_{0,0} \, \leq \omega^{(r)}_{\max} \, , \\
 -\omega^{(r)}_{\max} \leq \, &\hat{\omega}_{n_x, \, n_y} \leq \omega^{(r)}_{\max} \, , \forall \{n_x, \, n_y\} \neq \{0,0\} \, .
\end{align}
\end{subequations}

\section{Results} \label{results}

The STL computations are performed using the numeric PWE-based formulations presented in Section \ref{sound_insulation} for both the thickness-varying plates (PCs) and the constant-thickness plate with included mass-spring resonators (MMs). The results are computed using the material properties of aluminium (Young's modulus $E = 70$ GPa, Poisson's coefficient $\nu = 0.3$, and mass density $\rho = 2700$ kg/m$^3$). The mean plate thickness is chosen as $\overline{h} = 3$ mm and this is the constant value for analytical formulations. For these material properties and plate thickness, the critical frequency (Eq. (\ref{f_co})) is $4$ kHz. For the maximum frequency of $1$ kHz (well below the critical frequency), the maximum flexural wave speed (Eq. (\ref{flexural_wave_speed})) is $170$ m/s, which corresponds to a minimum wavelength $\lambda_{\min} = 170$ mm and a relation $\overline{h}/\lambda_{\min} \approx 1/57$. For the PC plate, the thickness remains in the range $[h_{\min}, \, h_{\max}] = [1, \, 5]$ mm. A square lattice of length $L = 40 h_{\max} = 200$ mm is also considered to ensure a surface that has a smooth variation and allows to disregard deviations in the fluid-structure interaction with respect to the plate mean thickness. For the double-leaf configuration, leaves are separated by $d = 1.5 h_{\max} = 7.5$ mm. The surrounding fluid is air at $20^o$C, with a sound speed $c_0 = 340$ m/s and mass density  $\rho_0 = 1.2$ kg/m$^3$. The PWE-based numerical method for computing the STL uses $n_G = 169$ plane waves ($N_{\max} = 6$). For the optimisation process, a total of $28$ parameters is used ($-6 \leq \{ n_x, \, n_y \} \leq 6$ with symmetric parameters, see Section \ref{optimization}).

\subsection{Comparison between analytical and numerical results}

\begin{figure}[!h]
    \centering
    \makebox[\textwidth]{
    \includegraphics[scale=0.8]{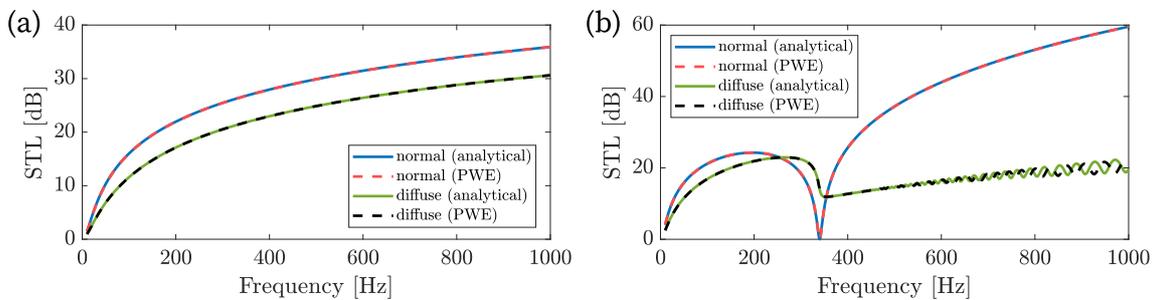}
    }
    \caption{Computed STLs for the (a) single-leaf and (b) double-leaf panels. The curves indicate the results for the normal and diffuse incidence cases, comparing both analytical and PWE approaches.}
    \label{validation}
\end{figure}

We start by comparing the STLs obtained for both normal ($\varphi=0^o$) and diffuse incidences ($0^o \leq \varphi \leq 78^o$ for the computation of Eq. (\ref{tau_d})) for both the analytical and PWE-based numerical solutions (Eqs. (\ref{set_u_wavy})). Figure \ref{validation} shows an excellent agreement between both. While no differences are noticed for the single-leaf case (figure \ref{validation}a),  a small deviation is noticed for the double-leaf diffuse case (figure \ref{validation}b), where the PWE approach presents slightly different STL values above $800$ Hz. An STL dip corresponding to the mass-air-mass resonance of the double-leaf system is also noticed at $340$ Hz.

The following section presents the optimisation results obtained for a one-octave frequency range centred at $350$ Hz (chosen close to the STL dip for the double-leaf initial case), i.e., from $248$ Hz to $495$ Hz.

\subsection{Optimisation results}

\begin{figure}[!h]
    \centering
    \makebox[\textwidth]{
    \includegraphics[scale=0.8]{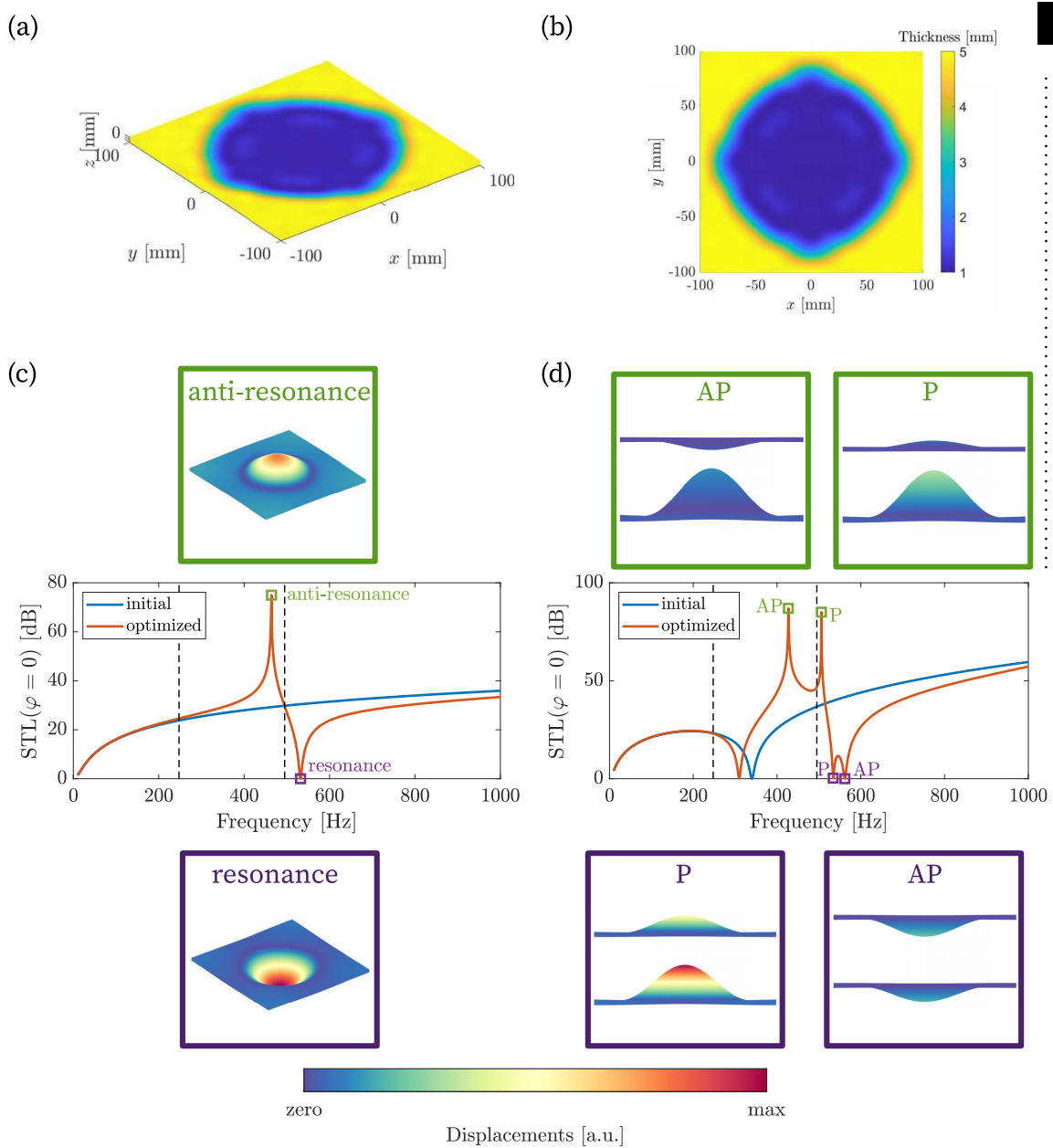}
    }
    \caption{Results for the PC panel considering normal incidence.
    (a,b) Optimised unit cell in the single-leaf case and (c) corresponding improvement in the STL within the target frequency range  due to an anti-resonance (green square), with a decrease of the STL outside the target frequency range due to resonance (purple square). (d) Same for the double-leaf case with optimised unit cell with anti-phase (AP) and in-phase (P) anti-resonances (green squares) and resonances  (purple squares).}
    \label{results_opt1_wavy_normal}
\end{figure}

The results for the PC case consider a fixed mean thickness of $h_0 = \overline{h} = 3$ mm. For the analysed frequency range, the obtained optimal thickness profile indicates a large area with the maximum thickness and a circular thickness reduction at the centre of the unit cell ($h_{\max} = 5$ mm and $h_{\min} = 1$ mm, respectively, see figures \ref{results_opt1_wavy_normal}a and \ref{results_opt1_wavy_normal}b). The computed normal incidence STL for the single-leaf case (figure \ref{results_opt1_wavy_normal}c) presents a large increase of the STL curve inside the frequency range of interest ($75.0$ dB at $464$ Hz), followed by a decrease outside this range ($0.01$ dB at $532$ Hz). The points of maximum and minimum correspond to the displacement profiles shown, for each case, using equivalent colour coding for the absolute displacement: an increase in the STL is achieved by a overall decrease in displacements of the unit cell (anti-resonance behaviour, represented in the green square), while a decrease in the STL curve is associated with an increase in the displacement of the region with the smallest thickness (resonance behaviour, represented in the purple square). It is also interesting to notice that no degradation is noticed before the upper limit of the optimisation frequency range. Very similar results are obtained when analysing other frequency ranges (see Supplementary Material).

For the double-leaf case, a similar result is obtained (figure \ref{results_opt1_wavy_normal}d), although, in this case, two anti-resonances ($87.0$ dB at $427$ Hz and $85.0$ dB at $506$ Hz) and two resonances ($0.29$ dB at $534$ Hz and $0.01$ dB at $562$ Hz) are present, which arises due to the combination of unit cell displacements in anti-phase (AP) and in-phase (P) combinations (shown with the same colour coding for proper comparison). Thus, an overall increase of performance is observed in the frequency range of interest. However, the STL dip due to the mass-air-mass resonance is still present.

\begin{figure}
 \centering
 \makebox[\textwidth]{
 \includegraphics[scale=0.8]{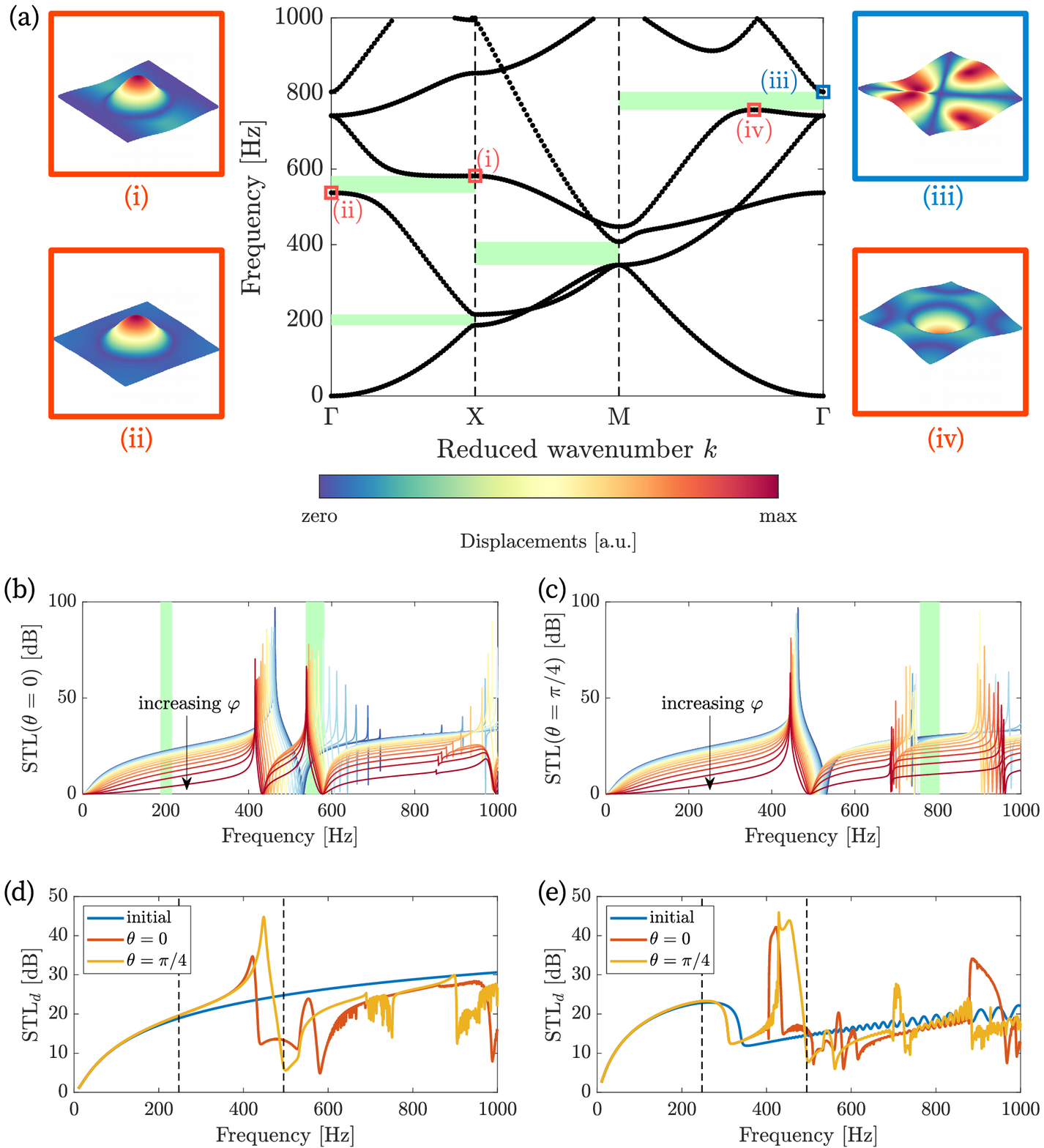}
 }
 \caption{Results for the PC panel under oblique and diffuse incidences.
 (a) Dispersion diagram for the optimised unit cell with partial band gaps (green regions) associated with wave modes (i), (ii), (iv) (red squares), and (iii) (blue square).
 STL considering oblique incidence angles for directions
 (b) $\theta=0$  and (c) $\theta=\pi/4$.
 STL for diffuse incidence for (d) single- and (e) double-leaf cases.}
 \label{results_opt1_wavy_oblique}
\end{figure}

However, one must recall that, for the normal incidence, the in-plane wavenumbers are zero (i.e., $k_x = k_y = 0$), and thus the characteristics observed in the STL for normal incidence are not necessarily associated with any particular in-plane wave propagation attenuation mechanisms. Thus, to correlate the STLs with the plate structural behaviour, we plot the dispersion relation for the PC plate along with some noticeable wave modes (figure \ref{results_opt1_wavy_oblique}a), which shows partial band gaps are opened due to Bragg scattering ($187.1$ Hz -- $215.3$ Hz and $537.4$ Hz -- $582.0$ Hz for the $\Gamma$X direction, $346.3$ Hz -- $407.5$ Hz for the XM region, and $756.7$ Hz -- $804.1$ Hz for the $\Gamma$M directions, respectively). Some wave modes associated with these band gaps may present the same displacement profiles as the resonance points of the normal incidence, as in the case of the wave modes indicated as (i), (ii), and (iv) (red squares), while other wave modes, such as the one indicated by (iii) (blue square), do not present the same displacement profile as a localized STL resonance. Thus, wave modes (i), (ii), and (iv) may be excited by acoustic impinging waves.

To confirm this correlation, the results for oblique ($\varphi \neq 0$) incident waves using the values $\varphi = \{ 0, \, 5^o, \, \dots, \, 85^o \}$ are plotted for the directions $\theta=0$ and $\theta = \pi/4$ and shown with the partial band gaps calculated for the direction $\Gamma$X and $\Gamma$M, respectively. For the $\Gamma$X direction ($\theta=0$, figure \ref{results_opt1_wavy_oblique}b), resonances are easily noticed both above and below the second band gap, which correspond to the wave modes (i) and (ii) excited at different frequencies. For the $\Gamma$M direction ($\theta = \pi/4$, figure \ref{results_opt1_wavy_oblique}c), this is noticeable for the wave mode (iv), just below the shown band gap. For frequencies immediately above the band gap the STL remains unaffected, since the associated wave mode (iii) is not excited by the incident acoustic wave. Thus, the immediate effect of the opened partial band gaps is to impede the formation of resonances in their interior, although anti-resonances may still be present. The consequence of these characteristics is noticed when analysing the diffuse STLs for the single- (figure \ref{results_opt1_wavy_oblique}d) and double-leaf cases (figure \ref{results_opt1_wavy_oblique}e). Although sharp peaks may be introduced in the diffuse STL for both cases, the target frequency range may also present ranges where degradation occurs.

For the MM case, the resonators are included keeping a constant unit cell mass. The plate thickness is now reduced to $h_{\min}$, while the mass corresponding to the thickness variation, $\Delta m = 216$ g, is added in the form of resonators, with resonating frequencies in the range $[\omega^{(r)}_{\min}, \, \omega^{(r)}_{\max}]/2\pi = [0.001, \, 10]$ kHz, thus allowing the resonators to vary over a wide frequency range. The optimisation process is performed for the cases of a single resonator and also multiple distributed resonators (chosen for a total number of $25$, in this case).

\begin{figure}[!h]
 \centering
 \makebox[\textwidth]{
 \includegraphics[scale=0.8]{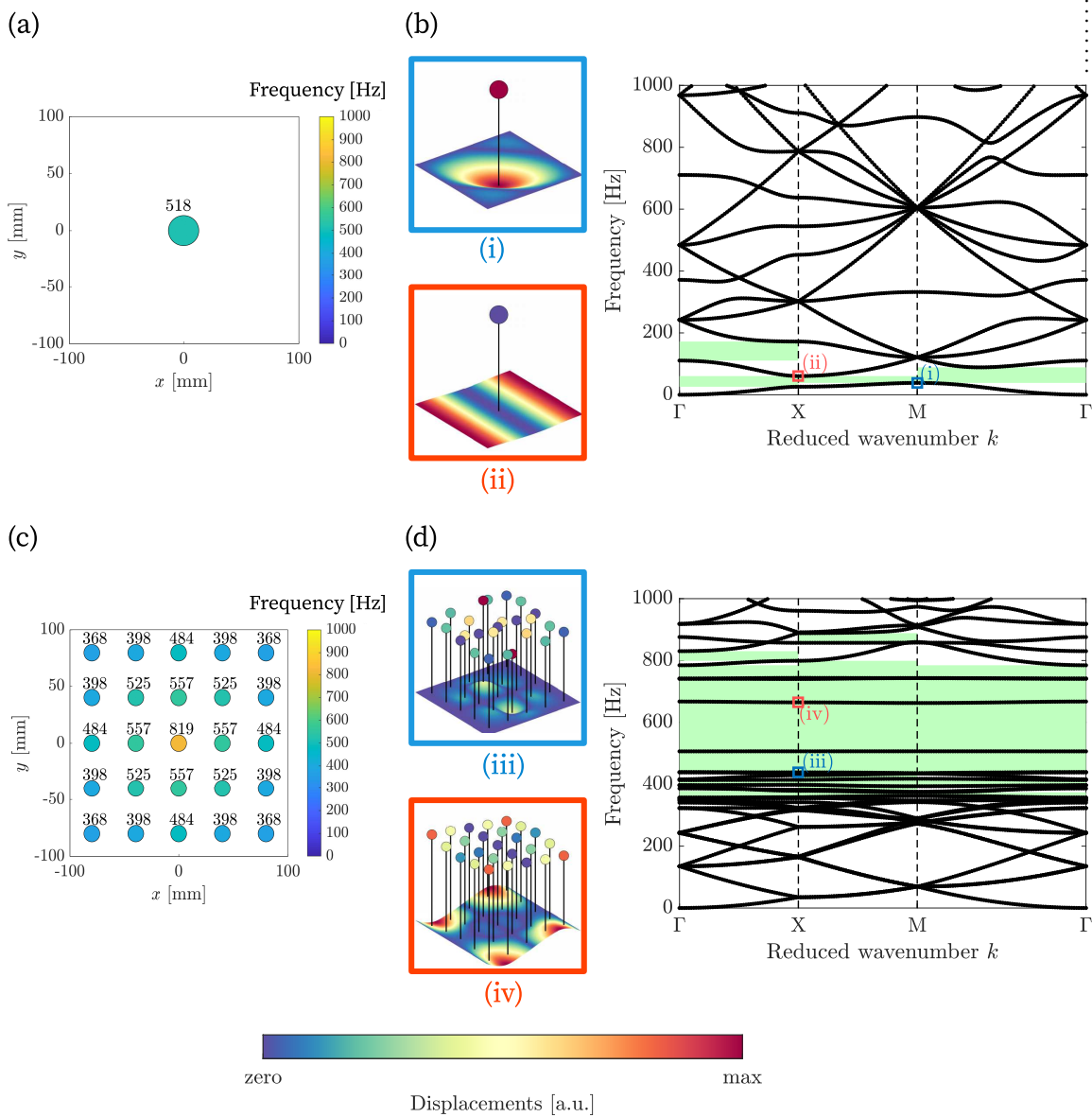}
 }
 \caption{Optimisation results for the MM panel.
    (a) Single-resonator result with 
    (b) dispersion diagram showing band gaps (green patches) and wave modes (i) and (ii) showing predominantly resonator (blue square) or plate displacements (red square).
    (c) Multiple-resonator result with
    (d) dispersion diagram showing
    several zero group velocity branches, opening large band gaps and wave modes (iii) and (iv).}
 \label{optimization_results_mm}
\end{figure}

We begin by analysing the resonator distribution obtained by each optimisation and their corresponding dispersion relations. For the single resonator case, a resonant frequency of $518$ Hz is indicated by the optimisation process (figure \ref{optimization_results_mm}a), which results in a dispersion diagram with a full band gap (i.e., for all propagation directions) in the $38.3$ Hz -- $60.5$ Hz frequency range (figure \ref{optimization_results_mm}b). The inclusion of a resonator with a large mass results in the flattening of the first flexural branch \cite{xiao2012flexural}, which is associated with a wave mode showing large displacements at the resonator (wave mode (i) at $38.3$ Hz, red square) and a wave mode with pronounced displacements at the plate (wave mode (ii) at $60.5$ Hz, blue square).

For the case of multiple resonators, the optimisation indicates one resonator with a higher frequency ($819$ Hz) and several resonators with lower frequencies ($368$ Hz -- $557$ Hz range, figure \ref{optimization_results_mm}c), with an appreciable superposition over the target optimisation frequency range. The resulting band diagram presents multiple flat bands (figure \ref{optimization_results_mm}d), which represent zero group velocity branches typical of locally resonant wave modes. Although this design presents a wide effective band gap (around $440$ Hz -- $785$ Hz) this may not necessarily translate into an effective STL gain, since, although some wave modes are associated with the displacement of resonators (wave mode (iii) at $438.7$ Hz, blue square), several wave modes are still associated with large plate displacements (e.g., wave mode (iv) at $664.9$ Hz, red square).

\begin{figure}[!hp]
 \centering
 \makebox[\textwidth]{
 \includegraphics[scale=0.8]{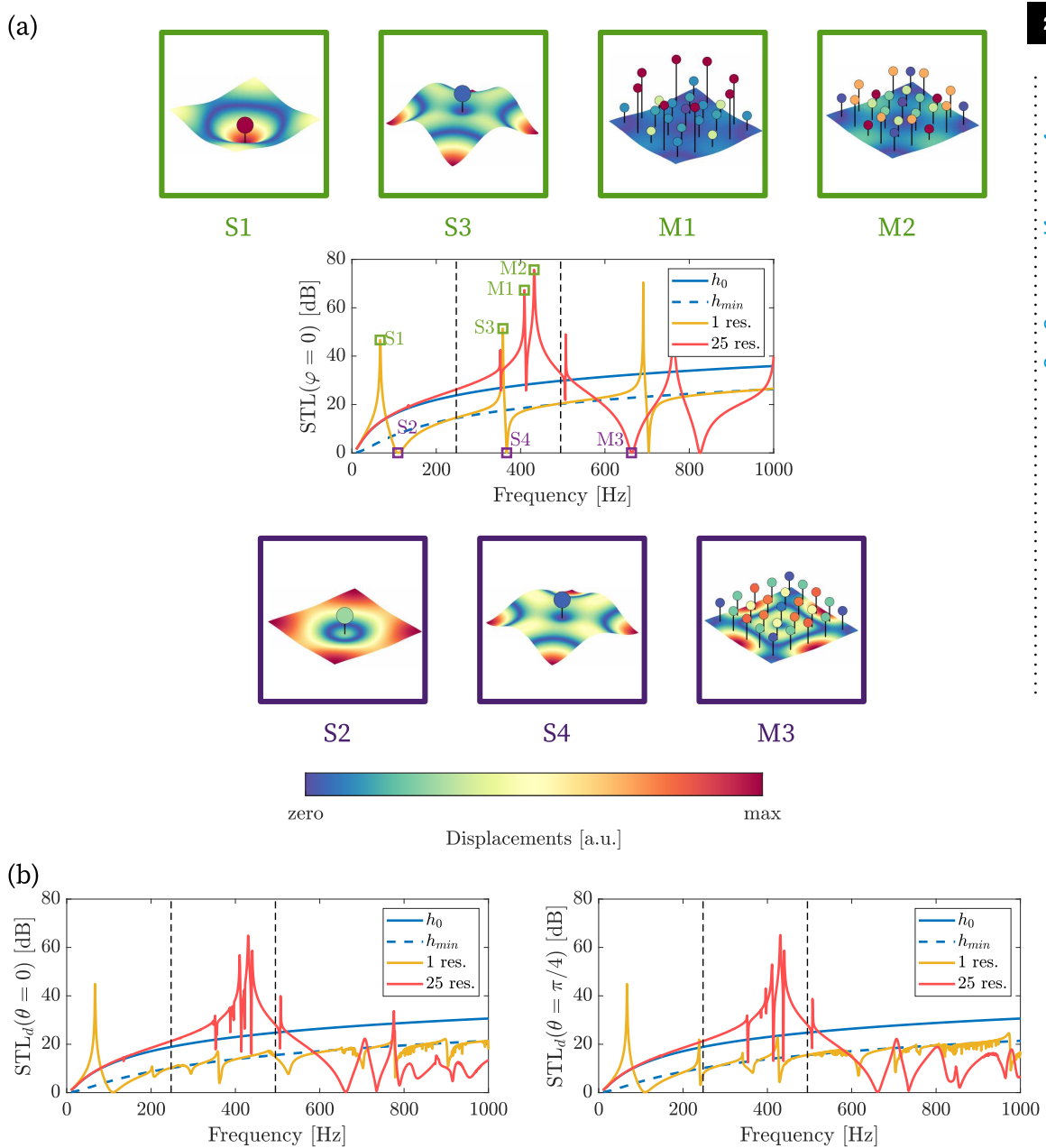}
 }
 \caption{Computed STLs for the single-leaf MM panel designs.
 (a) Normal-incidence case;
 improvements in the STLs are associated with localized resonator displacements (green squares, S1, S3, M1, and M2), while degradation occurs due to large plate displacements (purple squares, S2, S4, and M3), both related with the wave modes (figures \ref{optimization_results_mm}b and \ref{optimization_results_mm}d, (i)--(iv)).
 (b) Diffuse-incidences cases (directions $\theta=0$ and $\theta=\pi/4$) for minimum and maximum panel thickness, single-resonator, and multiple-resonator designs. The thinner plate ($h_{\min}$) is used for the attachment of resonators.}
 \label{stl_res_opt1_normal_single_results}
\end{figure}

The resulting STLs for the single-leaf under normal incidence are presented with the corresponding STLs for the constant-thickness plates with the mean thickness values of $\overline{h}$ (initial value) and $h_{\min}$ (on which the resonators are embedded) in figure \ref{stl_res_opt1_normal_single_results}a. The resulting STL is richer in dynamic behaviour when compared with the equivalent results for the PC. Examples of anti-resonance (green squares) and resonance behaviour (purple squares) are exemplified by S1 -- S4 (for the single resonator MM) and M1 -- M3 (for the multiple-resonator MM). Each displacement profile uses a colour scale normalized with respect to its maximum displacement for the purpose of comparing the plate and resonator displacements, while using the same out-of-plane displacement scaling factor.

The single resonator MM presents an STL similar to the plate with thickness $h_{\min}$, with deviations exemplified by the points labelled as S1 -- S4. Increases in this STL ($46.7$ dB at $67$ Hz for S1 and $51.4$ dB for $358$ Hz for S3, respectively) are associated with an overall reduction in the plate displacement, while decreases in the STL ($\approx 0$ dB at $110$ Hz for S2 and $0.03$ dB at $367$ Hz for S4) are associated with smaller resonator displacements and larger plate displacements. The resulting STL does not achieve an improvement over the desired optimisation frequency range. The multiple-resonator MM design is able to achieve a significantly higher STL. As in the PC design, no degradation is noticed in the STL until the upper edge of the optimisation frequency range.  Analogously to the single resonator MM design, the anti-resonances show a large displacement of the resonator masses and small plate displacements (e.g., $67.3$ dB at $409$ Hz for M1 and $75.7$ dB at $432$ Hz for M2), while for the resonances, this behaviour is the opposite (e.g., $\approx 0$ dB at $664$ Hz for M3).

For the multiple-resonator MM case, the wavelength-independence of the flat bands in the corresponding dispersion diagram (figure \ref{optimization_results_mm}d) also facilitate relating the wave modes with the anti-resonances and resonances: the anti-resonance M2 presents a frequency very similar to wave mode (iii) and the resonance M3 similar to wave mode (iv). Thus, although a wide band gap does exist and may be beneficial for structural applications, its effectiveness in improving the STL is conditioned to the shapes of the wave modes, which may be excited by impinging acoustic waves. Also, unlike the single resonator design, the use of several diverse resonators is able to achieve STL improvements in a broader frequency range, thus overcoming the shortcomings of the narrow frequency range influence of the resonant behaviour typically associated with single-frequency resonators.

The computed STLs using diffuse incidence for the single-leaf case are presented for the single- and multiple-resonator MM designs for the directions $\theta = 0$ (figure \ref{stl_res_opt1_normal_single_results}b) and $\theta = \pi/4$ (figure \ref{stl_res_opt1_normal_single_results}c). The resulting behaviour is very similar to the normal incidence case, although with smaller improvements in the STL. Also, it is interesting to notice that the wavelength independence of such locally-resonant based designs is insensitive to variations in the angle of incidence of the acoustic excitation, which implies an improved robustness for practical applications.

\begin{figure}[!h]
 \centering
 \makebox[\textwidth]{
 \includegraphics[scale=0.8]{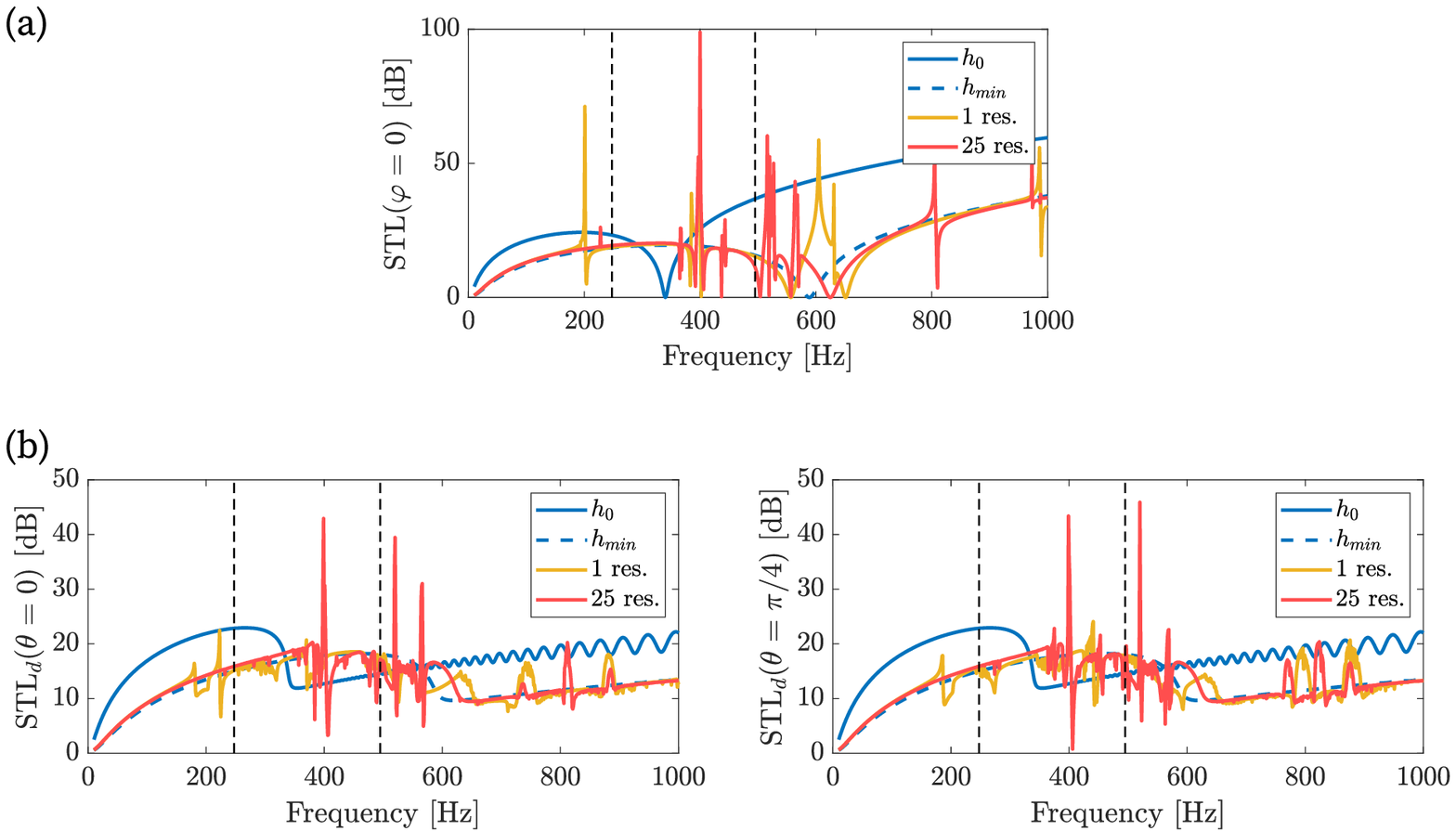}
 }
 \caption{Computed STLs for the double-leaf MM panel configuration.
    (a) Normal and (b) diffuse incidence cases
    for minimum and maximum panel thickness, single-resonator, and multiple-resonator designs.}
    \label{stl_res_opt1_other_results}
\end{figure}

The obtained results for both the single- and multiple-resonator double-leaf MM designs under normal incidence (figure \ref{stl_res_opt1_other_results}a) presents both STLs following the overall behaviour of the thinner plate, with the main characteristic of shifting the mass-air-mass resonant frequency to higher values, removing it from the optimisation frequency range. However, while the multi-resonator MM design introduces a single dip in the STL, the single-resonator design introduces two new dips at the vicinity of the previously existing one. Despite local deviations, the overall behaviour in the optimisation frequency range is nearly constant, presenting an overall degradation of the STL with respect to the thicker plate. A similar behaviour is observed for the double-leaf diffuse incidence case (figure \ref{stl_res_opt1_other_results}b, shown for $\theta=0$ and $\theta=\pi/4$), thus not presenting any justifiable gains over the original, thicker plate.

We have also performed an optimisation considering the possibility of changing both the thickness profile and resonator distribution, allowing the mean thickness to change while adding the corresponding mass reduction in the form of distributed resonators. The results yielded practically the same design as in the PC case (with deviations in the thickness parameters smaller than $1 \times 10^{-5}$) with zero resonator masses). This result is partially owed to the form of the integrand in the optimisation metric (see Eq. (\ref{optimization_metric})), which quickly converges to unity for sufficiently large improvements in the STL, thus favouring smaller STL improvements in wider frequency ranges (typical of PC designs) over large STL improvements in narrow frequency ranges (typical of MM designs).

\section{Conclusions} \label{conclusion}

Inspired by the locally resonant structures present in butterfly and moth wings and based on our previously work on thickness-varying plates for structural applications, we have investigated the utilization of these structures for acoustic insulation systems using both single- and double-leaf configurations.

With the proposed optimisation scheme, we obtained a PC plate with a thickness profile that presents an improvement in the STL under normal incidence, for both single- and double-leaf configurations, based on the anti-resonance behaviour of the unit cell. However, its dispersion diagram reveals that several wave modes may be excited by acoustic impinging waves outside of existing band gaps, thus degrading its performance for diffuse incidence cases.

The same optimisation procedure allows to obtain MMs constituted by plates with a reduced thickness and distributed resonators, while keeping the same unit cell mass. In this case, we obtained an optimised unit cell which produces wide band gaps associated with zero group velocity branches, with wave modes predominantly associated with either resonator or plate displacements. The use of multiple resonators with smaller masses achieves a superior STL performance when compared with a single resonator with a large mass, which is achieved by the excitation of the wave modes yielded by the different combinations of resonator displacements, thus obtaining improvements over a wider frequency range. The resulting STLs present similar behaviours for both normal and diffuse incidences, in contrast with the PC designs. For the double-leaf case, however, no real gains are achieved, since the resulting systems present a behaviour similar to that of a thinner plate.

In view of such results, it is clear that MM-based designs perform in a remarkably more robust manner for both normal and diffuse incidence cases for single-leaf configurations, due to their independence of the associated wavelength for the incident acoustic waves. These observations may also indicate why evolutionary pressure has led to the specialization of butterfly and moth wing structures in such a manner. The proposed PC-based designs, however, remain as interesting options when addressing normally-incident waves for both single- and double-leaf configurations.


\enlargethispage{20pt}

\dataccess{This article has no additional data.}

\aucontribute{ VFDP caried out the methodology, software, investigation, and writing of the original draft. NMP provided the computational resources, acquisition of funding, and performed the review of the manuscript. JRFA conceived and designed the study, performed the review of the manuscript, and overall supervision. All auhtors read and approved the manuscript.}

\competing{The authors declare that they have no competing interests.}

\funding{VFDP and NMP are supported by the EU H2020 FET Open ``Boheme'' grant No. 863179. JRFA thanks Conselho Nacional de Desenvolvimento Cient\'ifico e Tecnol\'ogico (CNPq), Brazil, grant 305293/2021-4, and Funda\c{c}\~ao de Amparo \`a Pesquisa do Estado de S\~ao Paulo (FAPESP), Brazil, grant 2018/15894-0.}



\bibliographystyle{rsta}
\bibliography{vibroacoustic_plate}


\end{document}